\documentclass{ws-ijmpb}

\usepackage{xcolor}
\usepackage[sort,compress]{cite}
\usepackage{subcaption}
\usepackage{graphicx}
\usepackage{caption}
\usepackage{amsmath}
\usepackage[bb=dsserif]{mathalpha}
\usepackage[verbose]{hyperref}
\hypersetup{colorlinks=false,allbordercolors=blue,pdfborderstyle={/S/U/W 1}}

\begin{document}

\markboth{Azzam S. Alzahrani and Victor M. Yakovenko}
{Loop currents in Haldane's model and in time-reversal-breaking superconductors}

%
\catchline{}{}{}{}{}
%

\title{Optical manifestations of loop currents in Haldane's model and in time-reversal-breaking superconductors}

\author{Azzam S. Alzahrani}

\address{Physics Department and Joint Quantum Institute\\
University of Maryland\\
College Park, MD 20742,
USA\\
azzamz@umd.edu}

\author{Victor M. Yakovenko}

\address{Physics Department and Joint Quantum Institute\\
University of Maryland\\
College Park, MD 20742,
USA\\
yakovenk@umd.edu}

\maketitle

\begin{history}
\received{Day Month Year}
\revised{Day Month Year}
\accepted{Day Month Year}
\published{Day Month Year}
\end{history}

\begin{abstract}
We present a theoretical study of optical manifestations of loop currents in Haldane's model and in time-reversal-breaking superconductors.  For Haldane's model, we calculate the expectation value of loop currents in terms of model parameters and relate it with the integrated optical spectral weight for the frequency-dependent ac Hall conductivity.  Thus, experimental measurements of the latter can provide information about the presence and magnitude of steady loop currents in the system.  Then we elaborate on loop currents in a chiral superconductor on the honeycomb lattice, studied earlier by Brydon \textit{et al.}\ (2019).  We demonstrate that a sharp optical absorption peak in the ac Hall conductivity originates from excitations between the lower and upper Dirac bands, activated by the time-reversal-breaking superconductivity.  The frequency of the peak is twice the energy difference between the Fermi level and the Dirac point.  The optical spectral weight of the peak is directly related to the magnitude of loop currents induced in the unit cells by the chiral superconducting pairing, in similarity to Haldane's model.

\end{abstract}

\keywords{loop currents; chiral superconductivity; polar Kerr effect; time-reversal symmetry breaking; optical Hall conductivity}

\section{Introduction}

Magnetism of materials, known since ancient times, has a quantum origin and primarily comes from the spin magnetic moments of the electrons.  The orbital magnetic moments of the electrons in atoms also contribute to their total magnetic moments. In addition to the intra-atomic magnetic moments, it is also possible, in principle, to have persistent loop currents circulating between different atoms within unit cells in a periodic crystal.  There is growing experimental evidence for such loop currents, as discussed by other articles in this special issue on Loop Currents.  As an example, recent measurements\cite{Suetsugu_2026} of the nuclear quadrupole resonance (NQR) spectra in a kagome metal indicate the presence of loop currents originating from the imaginary inter-site tunneling amplitudes, conceptualized as an imaginary charge density wave.  In our paper, we study theoretical models with loop currents. Although loop currents themselves are steady, i.e.,\ they are persistent dc currents at  zero frequency, we show that their presence has manifestations in frequency-dependent optical properties of the material.  Because loop currents break the time-reversal symmetry (TRS), their presence may produce a non-zero Hall conductivity in the absence of an external magnet field, i.e.,\ the anomalous Hall effect (AHE).  The AHE is often discussed as the dc effect at zero frequency\cite{Levchenko_2026}.  In contrast, we focus on the frequency-dependent optical AHE.  We show that the optical absorption spectrum of the Hall conductivity, i.e.,\ the antisymmetric time-reversal-odd part of the conductivity tensor, can provide useful information about steady loop currents in the system.  Alternatively, an optical vortex probe was recently proposed in Ref.~\refcite{Kato_2026} for the detection of loop currents in moir\'e materials.  So, optical tools can be useful for experimental studies of loop currents.

We consider two theoretical models with loop currents on the honeycomb lattice in two dimensions (2D): Haldane's model\cite{Haldane} in Sec.~\ref{Sec:Haldane} and the model of a chiral time-reversal-breaking superconductor\cite{Yakovenko} in Sec~\ref{Sec:SC}.  The presence of loop currents was pictorially indicated in Fig.~1 of Ref.~\refcite{Haldane} by arrows connecting the next-nearest-neighbor sites.  However, the expectation value of these loop currents was not calculated explicitly in this paper or elsewhere to the best of our knowledge.  We calculate the expectation value of loop currents in terms of model parameters and relate it with the integrated optical spectral weight of the frequency-dependent ac Hall conductivity.  We find that a non-zero expectation value of loop currents requires the presence of both real and imaginary parts of the next-nearest-neighbor tunneling amplitudes, whereas only the imaginary part is needed for a non-zero dc quantum Hall effect.  These results may be useful for interpreting the NQR measurements\cite{Suetsugu_2026} in a kagome metal.  Some optical experiments\cite{Xu_2022,Wang_2022} indicated a possible breaking of TRS in these materials, but other groups\cite{Kapitulnik_2023,Xia_2023,Xia_2024} argued against it.

In Ref.~\refcite{Yakovenko}, loop currents and ac AHE were studied for a tight-binding model on the honeycomb lattice with the time-reversal-breaking chiral superconducting pairing of the $d_{x^2-y^2}\pm id_{xy}$ symmetry.  Such chiral superconductivity was proposed theoretically in Ref.~\refcite{Chubukov_2012} for the highly-doped single-layer graphene, although this regime is currently not accessible experimentally.  A sharp optical peak in the calculated ac Hall conductivity was found in Fig.~6 of Ref.~\refcite{Yakovenko}.  However, the frequency of the peak and its origin were not discussed in this paper.  We demonstrate that the peak originates from electronic excitations between the lower and upper Dirac bands near the $K$ and $K'$ points in the Brillouin zone, activated by time-reversal-breaking superconductivity.  The frequency of the peak is twice the energy difference between the Fermi level and the Dirac point, which is typically much higher than the superconducting gap.  Similarly to Haldane's model, the optical spectral weight of the peak is directly related to the magnitude of loop currents induced in the unit cells by the chiral superconducting pairing.

The ac Hall conductivity results in the magneto-optical Kerr effect (MOKE), also known as the polar Kerr effect (PKE), which is a rotation of the polarization plane of linearly polarized light upon normal reflection for the surface of a material.  Its experimental detection was reported in a number of superconductors using an all-fiber Sagnac interferometer at wavelengths of 820 nm or 1550 nm, see Refs.~\refcite{Fried_2014,Paglione_2021} and references therein.  However, the signal is very weak at these high optical frequencies, so its observation requires a complicated calibration protocol and a custom-built apparatus\cite{Fried_2014}.  The signal is expected to be stronger at lower frequencies closer to a relevant electronic energy scale in a superconducting state.  For this reason, experimental measurements were recently performed at lower frequencies in the terahertz\cite{Armitage_2025} and sub-terahertz range\cite{Blumberg_2024}.  Moreover, these measurements\cite{Armitage_2025,Blumberg_2024} are spectroscopic in the sense that a signal is measured as a function of frequency in a certain range, rather than at a single frequency.  However, the interpretation of experimental observations requires a clear understanding of the spectral properties of ac Hall conductivity\cite{Armitage_2026}.  Our theoretical study contributes to this understanding.  We point out that the strongest signal in the Hall channel for clean superconductors can be obtained at the frequency of the absorption peak for interband transitions.  This is in contrast to superconductors with impurities\cite{Yakovenko_2009,Ostrovsky_2019}, where a peak is expected around the superconducting gap energy $2\Delta$.

\section{Haldane's model}
\label{Sec:Haldane}

\begin{figure}[b]
\centerline{\includegraphics[width=0.5\textwidth]{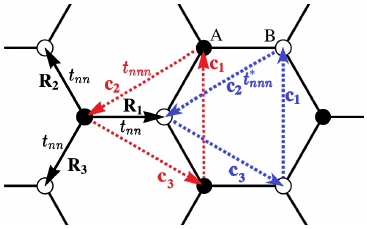}}
\caption{Haldane's model on the honeycomb lattice.}
\label{fig:Haldane}
\end{figure}

\subsection{The Hamiltonian}

Haldane proposed a tight-binding model of a Chern insulator on the honeycomb lattice in Ref.~\refcite{Haldane}.  As shown in Fig.~\ref{fig:Haldane}, electron hopping along the nearest-neighbor vectors $\bm{R}_1$, $\bm{R}_2$, and $\bm{R}_3$ has tunneling amplitudes $t_{\mathrm{nn}}$, while hopping along the next-nearest-neighbor vectors $\bm{c}_1$, $\bm{c}_2$ and $\bm{c}_3$ has  tunneling amplitudes $t_{\mathrm{nnn}}$.  The unit cell area $S_0$ can be expressed in terms of the nearest-neighbor spacing $R=|\bm{R}_j|$ and the next-nearest-neighbor spacing $c=|\bm{c}_j|=\sqrt{3}R$, where $j=1,2,3$,
\begin{equation}
S_0=\frac{S}{N}=\frac{3}{2}Rc=\frac{3\sqrt{3}}{2}R^2 ,
\label{eqn:unit_cell}
\end{equation}
where $S$ is the sample area and $N$ is the number of unit cells.

The TRS is broken by a periodic magnetic flux normal to the 2D plane, with zero net flux per unit cell. The nearest-neighbor hopping amplitude $t_{\mathrm{nn}}$ is real, whereas the next-nearest-neighbor hopping amplitude, $t_{\mathrm{nnn}}= t'_{\mathrm{nnn}}+it''_{\mathrm{nnn}}$, is complex due to magnetic flux. The Hamiltonian in real space reads
\begin{equation}
\begin{split}
\hat{H}_{\mathrm{Haldane}}=&\hat{H}_{\mathrm{nn}}+\hat{H}_{\mathrm{nnn}}, \\
\hat{H}_{\mathrm{nn}}=& \sum_{\bm{r}}\sum_{j=1,2,3}(t_{\mathrm{nn}}\hat{a}^\dagger_{\bm{r}}\hat{b}^{}_{\bm{r}+\bm{R}_j}+\text{h.c.}), \\
\hat{H}_{\mathrm{nnn}}=& \sum_{\bm{r}}\sum_{j=1,2,3}(t_{\mathrm{nnn}}\hat{a}^\dagger_{\bm{r}}\hat{a}^{}_{\bm{r}+\bm{c}_j}+t_{\mathrm{nnn}}\hat{b}^\dagger_{\bm{r}}\hat{b}^{}_{\bm{r}+\bm{c}_j}+\text{h.c.}),
\end{split}
\label{eqn:Hamiltonian_H}
\end{equation}
where we use the hat notation to denote the second-quantized operators $\hat{a}^\dagger_{\bm{r}} \ (\hat{a}^{}_{\bm{r}})$ and $\hat{b}^\dagger_{\bm{r}} \ (\hat{b}^{}_{\bm{r}})$ that create (annihilate) an electron at the position $\bm r$ on the A or B sublattice, respectively.  Throughout Sec.~\ref{Sec:Haldane}, we ignore the spin and assume that the chemical potential is zero: $\mu=0$, for simplicity.  So, the equations presented in Sec.~\ref{Sec:Haldane} effectively correspond to spineless fermions.

In momentum space, the Hamiltonian (\ref{eqn:Hamiltonian_H}) is expressed as
\begin{equation}
\begin{split}
\hat{H}_{\mathrm{Haldane}}=\sum_{\bm{k}}\hat{H}(\bm{k}),
\qquad
\hat{H}(\bm{k})=\hat{\Psi}^{\dagger}_{\bm{k}}H(\bm{k})\hat{\Psi}^{}_{\bm{k}}, 
\end{split}
\end{equation}
where
\begin{equation}
H(\bm{k})= \sum_{j=1,2,3} \begin{pmatrix}
  t_{\mathrm{nnn}} e^{i\bm{k}\cdot \bm{c}_j}+t^*_{\mathrm{nnn}}e^{-i\bm{k}\cdot \bm{c}_j}
  & t_{\mathrm{nn}} e^{i\bm{k}\cdot \bm{R}_j} \\ 
  t_{\mathrm{nn}} e^{-i\bm{k}\cdot \bm{R}_j}
  & t_{\mathrm{nnn}} e^{-i\bm{k}\cdot \bm{c}_j}+t^*_{\mathrm{nnn}}e^{i\bm{k}\cdot \bm{c}_j}
\end{pmatrix}.
\label{eqn:Hamiltonian_H_mat}
\end{equation}
The creation and annihilation operators are transformed as $\hat{\Psi}^{}_{\bm{k}}=(\hat{a}^{}_{\bm{k}},\hat{b}^{}_{\bm{k}})$, where $\hat{a}^\dagger_{\bm{k}} \ (\hat{a}^{}_{\bm{k}})$ and $\hat{b}^\dagger_{\bm{k}} \ (\hat{b}^{}_{\bm{k}})$ create (annihilate) an electron with wave vector $\bm{k}=(k_x,k_y)$ on the A and B sublattices, respectively. It is practical to parameterize the Hamiltonian (\ref{eqn:Hamiltonian_H_mat}) in terms of the Pauli matrices $\bm{\sigma}=(\sigma_1,\sigma_2,\sigma_3)$ and the unit matrix $\sigma_0$, 
\begin{align}
H(\bm{k})=&2t'_{\mathrm{nnn}}f_0(\bm{k})\sigma_0+\bm{w}(\bm{k})\cdot\bm{\sigma}, 
  \label{eqn:H-w} \\ 
\bm{w}(\bm{k})=& [t_{\mathrm{nn}}f_1(\bm{k}) ,\, t_{\mathrm{nn}}f_2(\bm{k}), \, 2t''_{\mathrm{nnn}}f_3(\bm{k})],
  \label{eqn:W}
\end{align}
where
\begin{equation}
\begin{split}
f_0(\bm{k})=& \sum_{j=1,2,3} \cos(\bm{k}\cdot \bm{c}_j)=2\cos\left( k_y\frac{c}{2}\right)\left[\cos\left(k_y\frac{c}{2}\right)+\cos\left(k_{x}\frac{\sqrt{3}c}{2}\right)\right]-1, \\
f_1(\bm{k})=& \sum_{j=1,2,3} \cos(\bm{k}\cdot \bm{R}_j)=2\cos\left( k_x\frac{R}{2}\right)\left[\cos\left(k_x\frac{R}{2}\right)+\cos\left(k_{y}\frac{\sqrt{3}R}{2}\right)\right]-1, \\
f_2(\bm{k})=& \sum_{j=1,2,3} \sin(\bm{k}\cdot \bm{R}_j)=2\sin\left( k_x\frac{R}{2}\right)\left[\cos\left(k_x\frac{R}{2}\right)-\cos\left(k_{y}\frac{\sqrt{3}R}{2}\right)\right], \\
f_3(\bm{k})=& \sum_{j=1,2,3} \sin(\bm{k}\cdot \bm{c}_j)=2\sin\left( k_y\frac{c}{2}\right)\left[\cos\left(k_y\frac{c}{2}\right)-\cos\left(k_{x}\frac{\sqrt{3}c}{2}\right)\right].
\end{split}
\label{eqn:f_function}
\end{equation}
The eigenenergies of Hamiltonian (\ref{eqn:H-w}) are
\begin{equation}
\varepsilon_{1,2}(\bm{k})=2t'_{\mathrm{nnn}}f_0(\bm{k})\mp w(\bm{k}),
\end{equation}
where $w(\bm{k})=|\bm{w}(\bm{k})|$. The corresponding Green's function is
\begin{equation}
G(\nu,\bm{k})=[i\nu \, \sigma_0-H(\bm{k})]^{-1}=\frac{[i\nu-2t'_{\mathrm{nnn}}f_0(\bm{k})]\sigma_0+\bm{w}(\bm{k})\cdot\bm{\sigma}}{(i\nu-\varepsilon_1)(i\nu-\varepsilon_2)},
\label{eqn:Green}
\end{equation}
where $\nu$ is the fermionic Matsubara frequency.

\subsection{Loop currents}

From the Hamiltonian in Eq.~(\ref{eqn:Hamiltonian_H}), consider the term for a bond connecting some sites $\bm{r}_1$ and $\bm{r}_2$:
\begin{equation}
\begin{split}
\hat{H}_{\bm{r}_1,\bm{r}_2}=t_{\bm{r}_1,\bm{r}_2}\hat{\psi}^\dagger_{\bm{r}_1}\hat{\psi}^{}_{\bm{r}_2}+\text{h.c.},
\end{split}
\end{equation}
where $\hat{\psi}_{\bm{r}}$ denotes $\hat{a}_{\bm{r}}$ or $\hat{b}_{\bm{r}}$, and $t_{\bm{r}_1,\bm{r}_2}$ is the tunneling amplitude between $\bm{r}_1$ and $\bm{r}_2$. Then the current operator $\hat{I}_{\bm{r}_1,\bm{r}_2}$=$-\hat{I}_{\bm{r}_2,\bm{r}_1}$ for the bond $(\bm{r}_1,\bm{r}_2)$ is defined as the time derivative of the charge on site $\bm{r}_1$ due to tunneling from site $\bm{r}_2$:
\begin{equation}
\begin{split}
\hat{I}_{\bm{r}_1,\bm{r}_2}=&\frac{d\hat{Q}}{dt}=e\frac{d}{dt}\left(\hat{\psi}^\dagger_{\bm{r}_1}\hat{\psi}^{}_{\bm{r}_1}\right)=-e\frac{d}{dt}\left(\hat{\psi}^\dagger_{\bm{r}_2}\hat{\psi}^{}_{\bm{r}_2}\right)\\=&i\frac{e}{\hbar}[\hat{\psi}^\dagger_{\bm{r}_1}\hat{\psi}^{}_{\bm{r}_1},\hat{H}_{\bm{r}_1,\bm{r}_2}]=i\frac{e}{\hbar}\left(t_{\bm{r}_1,\bm{r}_2}\hat{\psi}^\dagger_{\bm{r}_1}\hat{\psi}^{}_{\bm{r}_2}-\text{h.c.}\right),
\end{split}
\label{eqn:bond}
\end{equation}
where $e$ is the electron charge.

\begin{figure}[t]
\centering
    \includegraphics[width=0.45\textwidth]{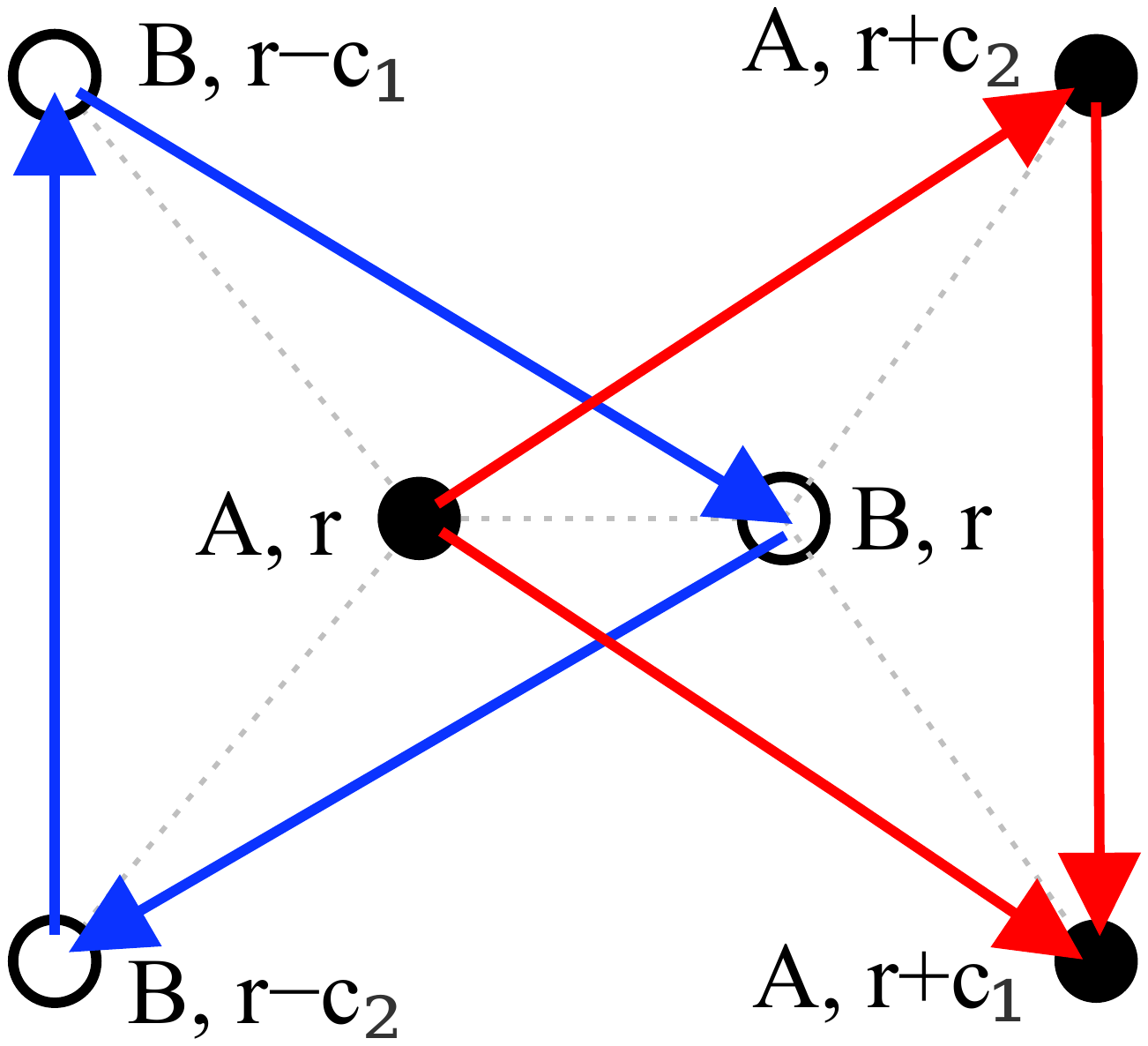}
\caption{Loop currents on the A and B sublattices.}
\label{fig:loop}
\end{figure}\noindent

The expectation value of the currents between nearest-neighbor sites on the honeycomb lattice vanishes due to a combination of rotational and translational symmetries.  In contrast, the expectation values of the currents on the next-nearest-neighbor bonds have the same magnitude that is generally nonzero, as illustrated in Fig.~\ref{fig:loop}. Thus, following Ref.~\refcite{Yakovenko}, we construct the loop current operator as
\begin{equation}
\hat{I}_{\mathrm{lc}}=i\frac{e}{\hbar}\frac{1}{6N}\sum_{\bm{r}}\sum_{j=1,2,3}\left(t_{\mathrm{nnn}}\hat{a}^\dagger_{\bm{r}}\hat{a}^{}_{\bm{r}+\bm{c}_j}+t_{\mathrm{nnn}}\hat{b}^\dagger_{\bm{r}}\hat{b}^{}_{\bm{r}-\bm{c}_j}-\text{h.c.} \right).
\label{eqn:loop_current_def}
\end{equation}
In momentum space, it can be written as
\begin{equation}
\begin{split}
\hat{I}_{\mathrm{lc}}=&i\frac{e}{\hbar}\frac{1}{6N}\sum_{\bm{k}}\sum_{j=1,2,3}\left(t_{\mathrm{nnn}}e^{i\bm{k}\cdot \bm{c}_j}\hat{a}^\dagger_{\bm{k}}\hat{a}^{}_{\bm{k}}+t_{\mathrm{nnn}}e^{-i\bm{k}\cdot \bm{c}_j}\hat{b}^\dagger_{\bm{k}}\hat{b}^{}_{\bm{k}}-\text{h.c.}\right) 
\\ 
=&-\frac{e}{\hbar}\frac{S_0}{3} \int \frac{d^2k}{(2\pi)^2}\hat{\Psi}^{\dagger}_{\bm{k}}I_{\mathrm{\mathrm{lc}}}(\bm{k})\hat{\Psi}^{}_{\bm{k}},
\end{split}
\label{eqn:loop_current}
\end{equation}
where the loop current matrix is
\begin{equation}
I_{\mathrm{\mathrm{lc}}}(\bm{k})=t''_{\mathrm{nnn}}f_0(\bm{k})\sigma_0+t'_{\mathrm{nnn}}f_3(\bm{k})\sigma_3.
\label{eqn:I_lc(k)}
\end{equation}
The expectation value of the loop current operator (\ref{eqn:loop_current}) in Haldane's model is
\begin{equation}
\overline{I_{\mathrm{lc}}}\equiv\langle \hat{I}_{\mathrm{lc}}\rangle
=-\frac{e}{\hbar}\frac{S_0}{3}\,T\sum_{\nu}\int \frac{d^2k}{(2\pi)^2}\mathrm{Tr}[I_{\mathrm{lc}}(\bm{k}) \, G(\nu,\bm{k})],
\end{equation}
where $T$ is the temperature.  Substituting $I_{\mathrm{\mathrm{lc}}}(\bm{k})$ from Eq.~(\ref{eqn:I_lc(k)}) and Green's function $G(\nu,\bm{k})$ from Eq.~(\ref{eqn:Green}), we find
\begin{equation}
\begin{split}
\overline{I_{\mathrm{lc}}}=&-\frac{2}{3}\frac{e}{\hbar}S_0 T\sum_{\nu}\int \frac{d^2k}{(2\pi)^2} \frac{t''_{\mathrm{nnn}}f_0(i\nu-2t'_{\mathrm{nnn}}f_0)+2t'_{\mathrm{nnn}}t''_{\mathrm{nnn}}f^2_3}{(i\nu-\varepsilon_1)(i\nu-\varepsilon_2)} \\
=&\frac{4}{3}\frac{e}{\hbar} S_0t'_{\mathrm{nnn}}t''_{\mathrm{nnn}}\int \frac{d^2k}{(2\pi)^2} \frac{f^2_0-f_3^2}{\varepsilon_2-\varepsilon_1} [F(\varepsilon_1)-F(\varepsilon_2)],
\end{split}
\label{<eqn:I_lc>}
\end{equation}
where $\varepsilon_2-\varepsilon_1=2w$, and $F$ is the Fermi-Dirac distribution.  The argument $\bm k$ is omitted in Eq.~(\ref{<eqn:I_lc>}) to shorten the notation.  At $T=0$, Eq.~(\ref{<eqn:I_lc>}) reduces to
\begin{equation}
\overline{I_{\mathrm{lc}}}=\frac{4}{3}\frac{e}{\hbar} S_0t'_{\mathrm{nnn}}t''_{\mathrm{nnn}}\int_{\rm\underline{BZ}}\frac{d^2k}{(2\pi)^2} \frac{f^2_0(\bm{k})-f_3^2(\bm{k})}{2w(\bm{k})},
\label{eqn:full_loop_current}
\end{equation}
where $\rm\underline{BZ}$ represents the region of the Brillouin zone that is singly occupied, i.e., where $\varepsilon_1(\bm{k})<0$, $F(\varepsilon_1)=1$ and $\varepsilon_2(\bm{k})>0$, $F(\varepsilon_2)=0$.

To our knowledge, Eq.~(\ref{eqn:full_loop_current}) is the first explicit calculation of the loop currents for Haldane's model, although it has been intuitively expected that those loop currents exist.  Note that a nonzero loop current $\overline{I_{\mathrm{lc}}}$ requires that both real $t_{\mathrm{nnn}}'$ and imaginary $t_{\mathrm{nnn}}''$ parts of the next-nearest-neighbor tunneling $t_{\mathrm{nnn}}$ be nonzero.

\subsection{The optical Hall spectral weight}

We introduce the optical Hall spectral weight
\begin{equation}
W_{\mathrm{Hall}}\equiv\frac{1}{2\pi}\int_0^\infty d\omega \, \omega\,\mathrm{Im}\,\sigma_{H}(\omega)
\label{eqn:weight}
\end{equation}
as the integral of the imaginary (absorptive) part of the ac Hall conductivity  $\sigma_{H}(\omega)$, which is defined as the antisymmetric part of the conductivity tensor $\sigma_{\alpha\beta}(\omega)$:
\begin{equation}
\sigma_{H}(\omega)=\frac{1}{2}\epsilon_{\alpha\beta}\sigma_{\alpha\beta}(\omega).
\end{equation}
Here $\alpha=x,y$ is the cartesian index and $\epsilon_{\alpha\beta}$ is the antisymmetric tensor in 2D, while $\omega$ is the ac frequency.  According to the magneto-optical sum rule derived in Ref.~\refcite{Kotliar} (also shown between Eqs.~(40) and (41) in Ref.~\refcite{Yakovenko}), the optical Hall spectral weight (\ref{eqn:weight}) is related to the expectation value of the commutator of current operators:
\begin{equation}
W_{\mathrm{Hall}}=-i\,\frac{\hbar}{8}
\frac{\left\langle[\hat{J}_{\alpha},\hat{J}_{\beta}]\right\rangle}{S}\epsilon_{\alpha\beta},
\label{eqn:sum_rule}
\end{equation}
where $S$ is the sample area. The current operator
\begin{equation}
\begin{split}
\hat{J}_{\alpha}=\frac{e}{\hbar}\sum_{\bm{k}}\hat{v}_\alpha(\bm{k})
\end{split}
\end{equation}
is defined in terms of the velocity operator 
$\hat{v}_{\alpha}(\bm{k})=\hat{\Psi}^{\dagger}_{\bm{k}} v^{}_{\alpha}(\bm{k})\hat{\Psi}^{}_{\bm{k}}$,
where the velocity matrix is
\begin{equation}
v_{\alpha}(\bm{k})=\frac{\partial H(\bm{k})}{\partial k_{\alpha}}=2t'_{\mathrm{nnn}}\frac{\partial f_0(\bm k)}{\partial k_{\alpha}}\sigma_0+\frac{\partial\bm w(\bm k)}{\partial k_{\alpha}}\cdot \bm{\sigma}.
\label{eqn:velocity_matrix}
\end{equation}
Using $\bm{w}(\bm{k})$ from Eq.~(\ref{eqn:W}), we find the following
\begin{equation}
\frac{\partial\bm{w}}{\partial k_{\alpha}}=\sum_{j=1,2,3} [-t_{\mathrm{nn}} \sin(\bm{k}\cdot \bm{R}_j)R_{j,\alpha},\, t_{\mathrm{nn}}\cos(\bm{k}\cdot \bm{R}_j)R_{j,\alpha}, \, 2t''_{\mathrm{nnn}} \cos(\bm{k}\cdot \bm{c}_j)c_{j,\alpha}].
\label{eqn:w_derivative}
\end{equation}
The commutator of the current operators is
\begin{align}
[\hat{J}_{\alpha},\hat{J}_{\beta}]=&\frac{e^2}{\hbar^2}\left[\sum_{\bm{k}'}\hat{\Psi}^{\dagger}_{\bm{k}'} v^{}_{\alpha}(\bm{k}')\hat{\Psi}^{}_{\bm{k}'},\sum_{\bm{k}''}\hat{\Psi}^{\dagger}_{\bm{k}''} v^{}_{\beta}(\bm{k}'')\hat{\Psi}^{}_{\bm{k}''}\right] 
\nonumber \\ 
=&\frac{e^2}{\hbar^2}\sum_{\bm{k}',\bm{k}''} [v^{}_{\alpha}(\bm{k}'),v^{}_{\beta}(\bm{k}'')]\,\hat{\Psi}^{\dagger}_{\bm{k}'} \{\hat{\Psi}^{}_{\bm{k}'}, \hat{\Psi}^{\dagger}_{\bm{k}''}\}\hat{\Psi}^{}_{\bm{k}''} 
\nonumber \\
=&\frac{e^2}{\hbar^2}\sum_{\bm{k}} [v^{}_{\alpha}(\bm{k}),v^{}_{\beta}(\bm{k})]
\,\hat{\Psi}^{\dagger}_{\bm{k}}\hat{\Psi}^{}_{\bm{k}},
\label{eqn:[JJ]}
\end{align}
where we used the fermionic anti-commutation relation $\{\hat{\Psi}^{}_{\bm{k}'}, \hat{\Psi}^{\dagger}_{\bm{k}''}\}=\delta_{\bm{k}',\bm{k}''}$.
Using Eq.~(\ref{eqn:velocity_matrix}), the commutator of the velocity matrices reads
\begin{equation}
[v_{\alpha}(\bm{k}),v_{\beta}(\bm{k})]=2i\left[\frac{\partial\bm{w}(\bm{k})}{\partial k_{\alpha}}\times\frac{\partial\bm{w}(\bm{k})}{\partial k_{\beta}}\right]\cdot\bm{\sigma}.
\label{eqn:commutator}
\end{equation}
The expectation value of the current commutator $\langle [\hat{J}_{\alpha},\hat{J}_{\beta}]\rangle$ per unit area in Eq.~(\ref{eqn:[JJ]}) can be obtained using Green's function from Eq.~(\ref{eqn:Green}),
\begin{align}
\frac{\langle [\hat{J}_{\alpha},\hat{J}_{\beta}]\rangle}{S}
=& 2i\frac{e^2}{\hbar^2}\int\frac{d^2k}{(2\pi)^2}\,T\sum_{\nu}
\mathrm{Tr}\left\{G(\nu,\bm{k})\,\bm{\sigma}\right\}\cdot\left[\frac{\partial\bm{w}(\bm{k})}{\partial k_{\alpha}}\times\frac{\partial\bm{w}(\bm{k})}{\partial k_{\beta}}\right]
\\
=& 4i\frac{e^2}{\hbar^2}\int\frac{d^2k}{(2\pi)^2}\,T\sum_{\nu}\frac{\bm{w}\cdot\left[\frac{\partial\bm{w}}{\partial k_{\alpha}}\times\frac{\partial\bm{w}}{\partial k_{\beta}}\right]}{(i\nu-\varepsilon_1)(i\nu-\varepsilon_2)}  
\\
=& 4i\epsilon_{\alpha\beta}\frac{e^2}{\hbar^2}\int\frac{d^2k}{(2\pi)^2}
\frac{K_{H}({\bm k})}{2w(\bm k)} 
\left\{F[\varepsilon_1(\bm k)]-F[\varepsilon_2(\bm k)]\right\}.
\label{eqn:velocities}
\end{align}
Here we introduced the antisymmetric velocity kernel
\begin{align}
K_{\alpha\beta}(\bm{k})
\equiv \epsilon_{\alpha\beta}K_{H}(\bm{k})
& \equiv\frac{1}{4i}\mathrm{Tr}\left\{H(\bm{k})\cdot\left[\frac{\partial H(\bm{k})}{\partial k_{\alpha}},\frac{\partial H(\bm{k})}{\partial k_{\beta}}\right]\right\} 
\label{eqn:K_H} \\
& =\bm{w}(\bm{k})\cdot\left[\frac{\partial\bm{w}(\bm{k})}{\partial k_{\alpha}}\times\frac{\partial\bm{w}(\bm{k})}{\partial k_{\beta}}\right].
\label{eqn:K_function}
\end{align}
The velocity kernel (\ref{eqn:K_function}) for Haldane's model is explicitly calculated in Appendix A and has the form
\begin{equation}
K_{H}(\bm{k})=\frac{2}{3}S_0t^2_{\mathrm{nn}}t''_{\mathrm{nnn}} [f^2_0(\bm{k})-f_3^2(\bm{k})].
\label{eqn:K_function_Full}
\end{equation}
Substituting Eq.~(\ref{eqn:K_function_Full}) into Eq.~(\ref{eqn:velocities}), we find the current commutator for Haldane's model
\begin{equation}
\begin{split}
\frac{\langle [\hat{J}_{\alpha},\hat{J}_{\beta}]\rangle}{S}=i\epsilon_{\alpha\beta}\frac{8}{3}S_0t^2_{\mathrm{nn}}t''_{\mathrm{nnn}}\frac{e^2}{\hbar^2}\int \frac{d^2k}{(2\pi)^2} \frac{f^2_0-f_3^2}{2w} [F(\varepsilon_1)-F(\varepsilon_2)].
\end{split}
\label{eqn:J_commutator}
\end{equation}
Comparing Eqs.~(\ref{<eqn:I_lc>}) and (\ref{eqn:full_loop_current}) with Eq.~(\ref{eqn:J_commutator}), we see that the loop current $\overline{I_{\mathrm{lc}}}$ can be expressed in terms of the current commutator, as well as the optical Hall spectral weight via Eq.~(\ref{eqn:sum_rule}):
\begin{equation}
\overline{I_{\mathrm{lc}}}=-\frac{i}{4}\frac{t'_{\mathrm{nnn}}}{t^2_{\mathrm{nn}}}\frac{\hbar}{e}\frac{\langle [\hat{J}_{\alpha},\hat{J}_{\beta}]\rangle}{S}\epsilon_{\alpha\beta}
= 2\frac{t'_{\mathrm{nnn}}}{t^2_{\mathrm{nn}}}\frac{1}{e}\, W_{\mathrm{Hall}}.
\label{eqn:currents_full}
\end{equation}
Thus, the optical Hall spectral weight $W_{\mathrm{Hall}}$ defined in Eq.~(\ref{eqn:weight}) is an indicator of the presence of loop currents and their magnitude.

\subsection{The ac Hall conductivity}

\begin{figure}[!t]
\centerline{\includegraphics[width=0.6\textwidth]{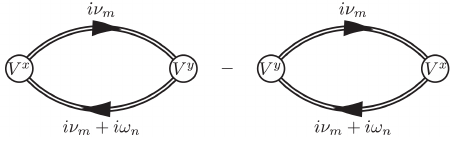}}
\vspace*{8pt}
\caption{Feynman diagram for the intrinsic Hall conductivity. The Matsubara frequencies $i\omega_n$ and $i\nu_m$ are the external bosonic and internal fermionic frequencies, respectively.}
\label{fig:Feynman}
\end{figure}

The intrinsic Hall conductivity is given by the Feynman diagram shown in Fig.~\ref{fig:Feynman},
\begin{equation}
\sigma_{H}(\omega)=i\frac{\epsilon_{\alpha\beta}}{2}
\frac{e^2}{\hbar^3\omega}\lim_{i\omega_n\to\omega+i0^+}\int \frac{d^2k}{(2\pi)^2} 
T\sum_\nu\mathrm{Tr}\{v_{\alpha}G(\nu)v_{\beta}G(\nu+\omega_n)\}.
\label{eqn:Hall_definition}
\end{equation}
Substituting Eqs.~(\ref{eqn:Green}) and (\ref{eqn:velocity_matrix}) into Eq.~(\ref{eqn:Hall_definition}) and evaluating the trace and the Matsubara sum, we express the antisymmetric tensor of the Hall conductivity in terms of the velocity kernel from Eq.~(\ref{eqn:K_function})
\begin{equation}
\sigma^H_{\alpha\beta}(\omega)=\epsilon_{\alpha\beta}\sigma_{H}(\omega)=\frac{e^2}{\hbar}\int \frac{d^2k}{(2\pi)^2} \frac{2K_{\alpha\beta}[F(\varepsilon_1)-F(\varepsilon_2)]}{w(2w-\omega-i0^+)(2w+\omega+i0^+)}.
\label{eqn:Hall_Haldane}
\end{equation}
Equation (\ref{eqn:Hall_Haldane}) is consistent with earlier derivations of ac Hall conductivity in Refs.~\refcite{Niu_2004,Yakovenko_2008,Yakovenko_2022}.  

In the limit $\omega=0$ at $T\to0$, dc Hall conductivity
\begin{equation}
\sigma^H_{\alpha\beta}(0)=\frac{e^2}{\hbar}\int_{\underline{BZ}} \frac{d^2k}{(2\pi)^2}\frac{K_{\alpha\beta}(\bm k)}{2w^3(\bm k)}
=\frac{e^2}{h}\int_{\underline{BZ}} \frac{d^2k}{4\pi}\,\Omega_{\alpha\beta}(\bm{k})
\label{eqn:DC_Hall}
\end{equation}
is expressed in terms of the Berry curvature tensor\cite{Berry}
\begin{equation}
\Omega_{\alpha\beta}(\bm{k})=\frac{K_{\alpha\beta}(\bm{k})}{w^3(\bm k)}=\bm{n}\cdot\left(\frac{\partial \bm{n}}{\partial k_{\alpha}}\times \frac{\partial \bm{n}}{\partial k_{\beta}}\right),
\label{eqn:Berry}
\end{equation}
where $\bm{n}(\bm k)=\bm{w}(\bm{k})/w(\bm{k})$. Substituting Eq.~(\ref{eqn:K_function_Full}) into Eq.~(\ref{eqn:Berry}) reproduces previous calculations\cite{Haldane_Berry} of the Berry curvature $\Omega_{\alpha\beta}(\bm{k})$ for Haldane's model.  When one band is occupied and another empty, the integral in Eq.~(\ref{eqn:DC_Hall}) is the degree of mapping and gives an integer Chern number. 

Using Eq.~(\ref{eqn:K_H}), the real and imaginary parts of the Hall conductivity from Eq.~(\ref{eqn:Hall_Haldane}) can be written as
\begin{align}
\mathrm{Re}\, \sigma_{H}(\omega)=&\frac{e^2}{h}\frac{2}{\pi}\int d^2k \, \frac{K_{H}(\bm{k})}{2w(\bm{k})\,[4w^2(\bm{k})-\omega^2]}\,
\{F[\varepsilon_1(\bm k)]-F[\varepsilon_2(\bm k)]\} 
\label{eqn:Real}\\
\mathrm{Im}\, \sigma_{H}(\omega)=&\frac{e^2}{h}\frac{1}{\omega^2}\int d^2k\, K_{H}(\bm{k})\,\delta[\omega-2w(\bm{k})]\,
\{F[\varepsilon_1(\bm k)]-F[\varepsilon_2(\bm k)]\}.
\label{eqn:Imag}
\end{align}
Equation (\ref{eqn:Imag}) matches Eq.~(20) in Ref.~\refcite{Yakovenko_2008}.  Note that Eqs.~(\ref{eqn:Real}) and (\ref{eqn:Imag}) satisfy the Kramers-Kronig relation (\ref{eqn:KK}).

Substituting Eq.~(\ref{eqn:Imag}) into Eq.~(\ref{eqn:weight}), we find that the optical Hall spectral weight is expressed as an integral of the current kernel $K_{H}(\bm{k})$:
\begin{equation}
W_{\mathrm{Hall}}=\frac{e^2}{\hbar} \int\frac{d^2k}{(2\pi)^2} \, 
\frac{K_{H}(\bm{k})}{2w(\bm{k})} 
\{F[\varepsilon_1(\bm k)]-F[\varepsilon_2(\bm k)]\}.
\label{eqn:weight_full}
\end{equation}
Comparing Eq.~(\ref{eqn:weight_full}) with Eq.~(\ref{eqn:velocities}), we confirm the magneto-optical sum rule in Eq.~(\ref{eqn:sum_rule}).

To conclude this section, we established that ac Hall conductivity $\sigma_{H}(\omega)$ and its optical spectral weight $W_{\mathrm{Hall}}$, as well as the expectation values of the loop currents $\overline{I_{\mathrm{lc}}}$ and the current commutator $\langle [\hat{J}_{\alpha},\hat{J}_{\beta}]\rangle$, can be expressed as integrals in the Brillouin zone of the velocity kernel $K_{H}(\bm{k})$.  Thus, the antisymmetric velocity kernel $K_{\alpha\beta}(\bm{k})$ introduced in Eq.~(\ref{eqn:K_H}) plays the central role and is similar, but more general than the Berry curvature tensor $\Omega_{\alpha\beta}(\bm{k})$, because the latter applies only to the dc case at $\omega=0$.

\section{Chiral superconductor}
\label{Sec:SC}

In Haldane's model discussed above, TRS is broken due to $t''_{\mathrm{nnn}}\neq0$.  In the following, we consider a model of superconductivity on the honeycomb lattice where $t''_{\mathrm{nnn}}=0$, so that TRS is not broken in the normal state. Instead, in this section, TRS is broken by the chiral $d_{x^2-y^2}\pm id_{xy}$ superconducting pairing.

\subsection{The model}

First, let us consider a general Bogoliubov-de Gennes (BdG) Hamiltonian of a spin-singlet superconductor with two sublattices A and B, perturbed with an electromagnetic vector potential $\bm{A}$
\begin{equation}
\begin{split}
\hat{H}(\bm{k},\bm{A})=\sum_{a,b,\bm{k}} 
&  \left[ \hat{\psi}^{\dagger}_{a,\uparrow,\bm{k}} H^{}_{0,ab}(\bm{k}-\frac{e}{\hbar c}\bm{A}) \hat{\psi}^{}_{b,\uparrow,\bm{k}}-\hat{\psi}^{}_{b,\downarrow,-\bm{k}} H_{0,ab}(-\bm{k}-\frac{e}{\hbar c}\bm{A}) \psi^{\dagger}_{a,\downarrow,-\bm{k}} \right.
\\ 
& \left. +\hat{\psi}^{}_{a,\downarrow,-\bm{k}} \Delta^{}_{ab}(\bm{k}) \hat{\psi}^{}_{b,\uparrow,\bm{k}}+\hat{\psi}^{\dagger}_{b,\uparrow,\bm{k}} \Delta^{*}_{ab}(\bm{k}) \hat{\psi}^{\dagger}_{a,\downarrow,-\bm{k}} \right],
\end{split}
\label{eqn:BdG}
\end{equation}
where $\hat{\psi}_{a,\uparrow,\bm{k}}$ annihilates an electron with spin $\uparrow$ and momentum $\bm{k}$ on the sublattice $a\in\{\rm A,B\}$.  We ignore the spin-orbit interaction for simplicity.  (The role of spin-orbit interaction was considered, e.g.,\ in Refs.~\refcite{Kallin_2017,Brydon,Niu_2026}.)  The effect of the vector potential $\bm{A}$ will be discussed in Sec. 3.3, so we set $\bm{A}=0$ for now. Introducing the 4-component operator $\hat{\Psi}^{}_{\bm{k}}=(\hat{\psi}^{}_{A,\uparrow,\bm{k}},\hat{\psi}^{}_{B,\uparrow,\bm{k}},\hat{\psi}^{\dagger}_{A,\downarrow,-\bm{k}},\hat{\psi}^{\dagger}_{B,\downarrow,-\bm{k}})$, the unperturbed BdG Hamiltonian at $\bm{A}=0$ can be written in terms of a $4\times4$ matrix $\check{H}(\bm k)$ as
\begin{equation}
\begin{split}
\hat{H}(\bm{k})= \sum_{\bm k} \hat{\Psi}^{\dagger}_{\bm{k}} \check{H}^{}(\bm{k})  \hat{\Psi}^{}_{\bm{k}}.
\end{split}
\end{equation}
We use the inverted hat notation to denote the $4\times4$ BdG matrix $\check{H}(\bm{k})$, which, in turn, is represented in terms of $2\times2$ matrices $H_0(\bm{k})$ and $\Delta(\bm{k})$
\begin{equation}
    \check{H}^{}(\bm{k})=\begin{pmatrix} \ H_{0}(\bm{k}) & \Delta(\bm{k}) \  \\  \ \Delta^\dagger(\bm{k}) & -H_{0}^T(-\bm{k})\ \end{pmatrix}. \quad 
\end{equation}
Here, the superscript $T$ (not to be confused with temperature) represents matrix transposition with respect to the sublattice indices $a$ and $b$. The normal-state Hamiltonian $H_0(\bm{k})=H_0^{\dagger}(\bm{k})$ is Hermitian, and we assume that it is invariant under time-reversal: $H_0(\bm{k})=H_0^{*}(-\bm{k})$. Combining these two properties, we find that $H^{T}_{0}(-\bm{k})=H^{}_{0}(\bm{k})$. Then, the $4\times4$ BdG Hamiltonian becomes
\begin{equation}
    \check{H}^{}(\bm{k})=\begin{pmatrix} \ H_{0}(\bm{k}) & \Delta(\bm{k}) \  \\  \ \Delta^\dagger(\bm{k}) & -H_{0}(\bm{k})\ \end{pmatrix}. \quad 
\label{eqn:BdG_general}
\end{equation}
In contrast, the superconducting pairing matrix is generally not Hermitian: $\Delta(\bm{k})\neq \Delta^{\dagger}(\bm{k})$, and it may break TRS. Equation (\ref{eqn:BdG_general}) represents the most general two-band superconductor where $H_0(\bm{k})$ has TRS, but $\Delta(\bm{k})$ may break it.

\begin{figure}[!t]
\centerline{\includegraphics[width=0.5\textwidth]{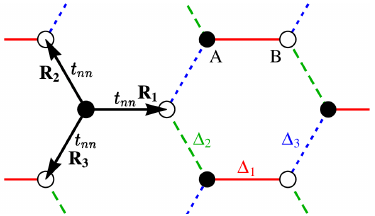}}
\caption{The model of a chiral superconductor on the honeycomb lattice.}
\label{fig:SC}
\end{figure}

Now, we apply this general formula specifically to the model of superconducting pairing on the honeycomb lattice that was studied in Ref.~\refcite{Yakovenko}, as illustrated in Fig.~\ref{fig:SC}. The normal-state Hamiltonian $H_0(\bm{k})$ is very similar to the Hamiltonian in Eq.~(\ref{eqn:Hamiltonian_H_mat}) for the Haldane model:
\begin{equation}
    H_{0}(\bm{k})=\begin{pmatrix} \ -\mu & t_{\mathrm{nn}}\sum_j e^{i\bm{k}\cdot \bm{R}_j} \  \\  \ t_{\mathrm{nn}}\sum_j e^{-i\bm{k}\cdot \bm{R}_j} & -\mu\ \end{pmatrix},
\label{eqn:Hamiltonian}
\end{equation}
where we introduced the chemical potential $\mu\neq0$. However, in contrast to Haldane's model, we assume that $H_0(\bm{k})$ does not break TRS, so $t''_{\mathrm{nnn}}=0$ is taken to be zero. Moreover, we assume that $t'_{\mathrm{nnn}}$ is very small, so it can be neglected  everywhere except for the definition of the loop current operator $\hat{I}_{\mathrm{lc}}$ discussed in Sec.~\ref{Sec:SC-loop-currents}.

\begin{figure}[!t]
\centering
\begin{subfigure}[t]{0.28\textwidth}
    \includegraphics[width=\textwidth]{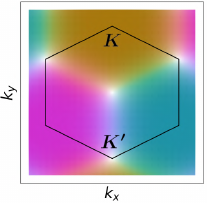}
    \caption{$\Delta_{AB}(\bm{k})$}
\end{subfigure}
~
\begin{subfigure}[t]{0.28\textwidth}
    \includegraphics[width=\textwidth]{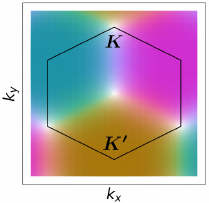}
    \caption{$\Delta_{BA}(\bm{k})$}
\end{subfigure}
~
\begin{subfigure}[t]{0.04\textwidth}
    \includegraphics[width=\textwidth]{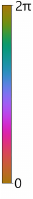}
\end{subfigure}
\caption{The inter-sublattice superconducting pairing amplitudes $\Delta_{AB}(\bm{k})$ in panel (a) and $\Delta_{BA}(\bm{k})$ in panel (b) over the Brillouin zone. The phase and magnitude are represented by color and saturation, respectively. The amplitude $\Delta_{AB}(K')=0$ vanishes at the $K'$ point, whereas $\Delta_{BA}(K)=0$ vanishes at the $K$ point.}
\label{fig:Pairing}
\end{figure}

As in Ref.~\refcite{Yakovenko}, we assume that superconducting pairing occurs on the bonds between the nearest-neighbor sites $j=1,2,3$ with the phases $\phi_{j}\in\{0,\frac{2\pi}{3},\frac{4\pi}{3}\}$, as shown in Fig.~\ref{fig:SC}. This sequence of  phases corresponds to the spin-singlet $d_{x^2-y^2}+id_{xy}$ pairing on the honeycomb lattice. Thus, the $2\times2$ matrix of the superconducting pairing is
\begin{equation}
    \Delta({\bm{k}})\equiv\begin{pmatrix} \ 0&  \Delta_{AB}({\bm{k}}) \ 
    \\ \  \Delta_{BA}({\bm{k}}) & 0 \ \end{pmatrix}
 =\begin{pmatrix} \ 0&  \Delta\sum_j e^{i\phi_j}e^{i\bm{k}\cdot \bm{R}_j} \ 
 \\ \  \Delta\sum_j e^{i\phi_j}e^{-i\bm{k}\cdot \bm{R}_j} & 0 \ \end{pmatrix}.
\label{eqn:Pairing}
\end{equation}
The pairing potentials $\Delta^{}_{AB}(\bm{k})\neq\Delta^{\dagger}_{BA}(\bm{k})$ are shown in Fig.~\ref{fig:Pairing}. Note that $\Delta^{}_{BA}(\bm k)$ has a vortex in momentum space around the $K$ point in the Brillouin zone and vanishes at this point: $\Delta^{}_{BA}(K)=0$.  In contrast, $\Delta^{}_{AB}(\bm k)$ has a vortex around the point $K'$ and vanishes at that point: $\Delta^{}_{AB}(K')=0$.

Green's function $\check{G}_{}(\nu,\bm{k})=[i\nu \, \check{\mathbb{1}}-\check{H}_{}(\bm{k})]^{-1}$ corresponding to Eq.~(\ref{eqn:Hamiltonian}) and (\ref{eqn:Pairing}) is derived in Appendix B. It is also shown in Appendix B that the energy eigenvalues of the BdG Hamiltonian come in pairs $\pm E_1(\bm{k})$ and $\pm E_2(\bm{k})$.

\subsection{Loop currents}
\label{Sec:SC-loop-currents}

\begin{figure}[!b]
\centering
\begin{subfigure}[t]{0.5\textwidth}
    \includegraphics[width=\textwidth]{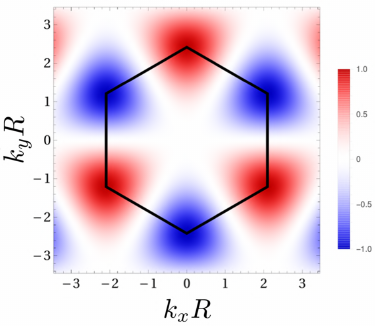}
\end{subfigure}
\caption{A 2D graph of the function $f_3(\bm{k})$ from Eq.~(\ref{eqn:f_function}) over the Brillouin zone. The function $f_3(\bm{k})$ appears in the loop current $\check{I}_{\mathrm{lc}}(\bm{k})$, Eq.~(\ref{eqn:loop_tensor}), in the superconducting sublattice polarization $\Xi(\bm{k})$, Eq.~(\ref{eqn:polarization}), and in the velocity commutator $\Upsilon(\bm k)$, Eq.~(\ref{eqn:Upsilon-f3}).}
\label{fig:Berry}
\end{figure}

Equation (\ref{eqn:bond}) for the current of a bond can be generalized to include spin:
\begin{equation}
\begin{split}
\hat{I}_{\bm{r}_1,\bm{r}_2}=&\hat{I}^{\uparrow}_{\bm{r}_1,\bm{r}_2}+\hat{I}^{\downarrow}_{\bm{r}_1,\bm{r}_2},\\
\hat{I}^{\uparrow}_{\bm{r}_1,\bm{r}_2}=&i\frac{e}{\hbar}\left(t_{\bm{r}_1,\bm{r}_2}\hat{\psi}^\dagger_{\bm{r}_1\uparrow}\hat{\psi}^{}_{\bm{r}_2\uparrow}-\text{h.c.}\right), \\
\hat{I}^{\downarrow}_{\bm{r}_1,\bm{r}_2}=&i\frac{e}{\hbar}\left(t_{\bm{r}_1,\bm{r}_2}\hat{\psi}^\dagger_{\bm{r}_1\downarrow}\hat{\psi}^{}_{\bm{r}_2\downarrow}-\text{h.c.}\right) \\
=&i\frac{e}{\hbar}\left(t^{*}_{\bm{r}_1,\bm{r}_2}\hat{\psi}^{}_{\bm{r}_1\downarrow}\hat{\psi}^\dagger_{\bm{r}_2\downarrow}-\text{h.c.}\right).
\end{split}
\end{equation}
Similarly to Eq.~(\ref{eqn:loop_current_def}), we construct the loop current operator in the extended Nambu basis as
\begin{equation}
\begin{split}
\hat{I}_{\mathrm{lc}}=&\hat{I}^{\uparrow}_{\mathrm{lc}}+\hat{I}^{\downarrow}_{\mathrm{lc}},\\
\hat{I}^{\uparrow}_{\mathrm{lc}}=&i\frac{e}{\hbar}\frac{1}{6N}\sum_{\bm{r}}\sum_{j=1,2,3}\left(t_{\mathrm{nnn}}\hat{\psi}^{\dagger}_{\bm{r}_A,\uparrow}\hat{\psi}^{}_{\bm{r}_A+\bm{c}_j,\uparrow}+t_{\mathrm{nnn}}\hat{\psi}^{\dagger}_{\bm{r}_B,\uparrow}\hat{\psi}^{}_{\bm{r}_B-\bm{c}_j,\uparrow}-\text{h.c.} \right),\\
\hat{I}^{\downarrow}_{\mathrm{lc}}=&i\frac{e}{\hbar}\frac{1}{6N}\sum_{\bm{r}}\sum_{j=1,2,3}\left(t^{*}_{\mathrm{nnn}}\hat{\psi}^{}_{\bm{r}_A,\downarrow}\hat{\psi}^{\dagger}_{\bm{r}_A+\bm{c}_j,\downarrow}+t^{*}_{\mathrm{nnn}}\hat{\psi}^{}_{\bm{r}_B,\downarrow}\hat{\psi}^{\dagger}_{\bm{r}_B-\bm{c}_j,\downarrow}-\text{h.c.} \right),
\end{split}
\end{equation}
In momentum space, it can be written as
\begin{equation}
\begin{split}
\hat{I}^{\uparrow}_{\mathrm{lc}}=&i\frac{e}{\hbar}\frac{1}{6N}\sum_{\bm{k}}\sum_{j=1,2,3}\left(t_{\mathrm{nnn}}e^{i\bm{k}\cdot \bm{c}_j}\hat{\psi}^{\dagger}_{A,\uparrow,\bm{k}}\hat{\psi}^{}_{A,\uparrow,\bm{k}}+t_{\mathrm{nnn}}e^{-i\bm{k}\cdot \bm{c}_j}\hat{\psi}^{\dagger}_{B,\uparrow,\bm{k}}\hat{\psi}^{}_{B,\uparrow,\bm{k}}-\text{h.c.}\right), \\
\hat{I}^{\downarrow}_{\mathrm{lc}}=&i\frac{e}{\hbar}\frac{1}{6N}\sum_{\bm{k}}\sum_{j=1,2,3}\left(t^{*}_{\mathrm{nnn}}e^{i\bm{k}\cdot \bm{c}_j}\hat{\psi}^{}_{A,\downarrow,-\bm{k}}\hat{\psi}^{\dagger}_{A,\downarrow,-\bm{k}}+t^{*}_{\mathrm{nnn}}e^{-i\bm{k}\cdot \bm{c}_j}\hat{\psi}^{}_{B,\downarrow,-\bm{k}}\hat{\psi}^{\dagger}_{B,\downarrow,-\bm{k}}-\text{h.c.}\right).
\end{split} \nonumber
\end{equation}
Taking into account that $t_{\mathrm{nnn}}=t'_{\mathrm{nnn}}$ because $t''_{\mathrm{nnn}}=0$, the loop current operator is
\begin{equation}
\hat{I}_{\mathrm{lc}}=-\frac{e}{\hbar}\frac{S_0}{3}\int \frac{d^2k}{(2\pi)^2}\hat{\Psi}^{\dagger}_{\bm{k}} \check{I}_{\mathrm{lc}}^{}(\bm{k})  \hat{\Psi}^{}_{\bm{k}},
\label{eqn:SC_lc_operator}
\end{equation}
with the $4\times4$ matrix
\begin{equation}
\check{I}_{\mathrm{lc}}^{}(\bm{k})=\begin{pmatrix} \ t'_{\mathrm{nnn}}f_3(\bm{k})\sigma_3&  0 \ \\ \  0 & t'_{\mathrm{nnn}}f_3(\bm{k})\sigma_3 \ \end{pmatrix}=t'_{\mathrm{nnn}}f_3(\bm{k})\tau_0\otimes\sigma_3.
\label{eqn:loop_tensor}
\end{equation}
Here, the unit matrix $\tau_0$ acts in the Nambu space, whereas $\sigma_3$ acts in the sublattice space. The function $f_{3}(\bm{k})$ is defined in Eq.~(\ref{eqn:f_function}) and is shown in Fig.~\ref{fig:Berry}.

\noindent

The expectation value of the loop current operator, Eq.~(\ref{eqn:SC_lc_operator}), can be obtained using Green's function $\check G$
\begin{equation}
\overline{I_{\mathrm{lc}}}\equiv\langle \hat{I}_{\mathrm{lc}}\rangle
=-\frac{S_0}{3}\frac{e}{\hbar}\,T\sum_{\nu}\int \frac{d^2k}{(2\pi)^2}
\mathrm{Tr}_{\tau\sigma}
\left\{\check{I}_{\mathrm{lc}}(\bm{k}) \, \check G_{}(\nu,\bm{k})\right\}.
\label{eqn:loop_current_expectation}
\end{equation}
Here, the subscript of the trace operator $\mathrm{Tr}_{\tau\sigma}$ indicates whether it is taken in the space of $\tau$, $\sigma$, or both.  Given that $\check I_{\mathrm{lc}}$ has the structure of $\tau_0\otimes\sigma_3$ in Eq.~(\ref{eqn:loop_tensor}), only the term in Eq.~(\ref{eqn:App_Green}) for $\check G$ with the matching structure, $G_{0,3}$ in Eq.~(\ref{eqn:All_Green}), contributes to Eq.~(\ref{eqn:loop_current_expectation}).  This term is expressed in terms of the superconducting sublattice polarization $\Xi(\bm{k})$
\begin{equation}
\Xi(\bm{k}) \equiv 
\mathrm{Tr}_\sigma\{\Delta^\dagger(\bm{k})\sigma_3 \Delta^{}(\bm{k})\}
 = |\Delta_{AB}({\bm{k}})|^2 - |\Delta_{BA}({\bm{k}})|^2,
\label{eqn:Xi_definition}
\end{equation}
which is related to the commutator of the pairing potentials, called the TROB in Ref.~\refcite{Yakovenko},
\begin{equation}
[\Delta(\bm{k}),\Delta^\dagger(\bm{k})] = \sigma_3 \, \Xi(\bm{k}).
\label{eqn:TROB}
\end{equation}
Using Eq.~(16) from Ref.~\refcite{Yakovenko}, we find in our notation that
\begin{equation}
\Xi(\bm{k}) = 2\sqrt{3}|\Delta|^2f_3(\bm{k})
\label{eqn:polarization}
\end{equation}
contains the same function $f_3(\bm{k})$ as shown in Fig.~\ref{fig:Berry}.  The pattern in Fig.~\ref{fig:Berry} is consistent with the difference of $|\Delta_{AB}({\bm{k}})|^2$ and $|\Delta_{BA}({\bm{k}})|^2$ shown in Panels (a) and (b) in Fig.~\ref{fig:Pairing}.

 Substituting Eq.~(\ref{eqn:loop_tensor}) and the term $G_{0,3}$ from Eq.~(\ref{eqn:All_Green}) into Eq.~(\ref{eqn:loop_current_expectation}) and evaluating the trace, we find
\begin{equation}
\overline{I_{\mathrm{lc}}}=-t'_{\mathrm{nnn}}\frac{e}{\hbar}\frac{2}{3}S_0
T\sum_{\nu}\int \frac{d^2k}{(2\pi)^2}  
\frac{\mu\,\Xi(\bm{k})\,f_3(\bm{k})}
{[\hbar^2\nu^2+E^2_1(\bm{k})]\,[\hbar^2\nu^2+E^2_2(\bm{k})]},
\label{eqn:loop_PRX}
\end{equation}
where $E_{1(2)}(\bm k)$ is the energy of the first (second) eigenstate of the BdG Hamiltonian. Equation (\ref{eqn:loop_PRX}) matches Eqs.~(26) and (30) in Ref.~\refcite{Yakovenko}. 

The Matsubara sum in Eq.~(\ref{eqn:loop_PRX}) is evaluated as
\begin{equation}
\begin{split}
&\sum_{\nu} \frac{1}{(\hbar^2\nu^2+E^2_{1})(\hbar^2\nu^2+E^2_{2})}\equiv\sum_{\nu} h({\nu}) \\
&=\sum_{Z_n} i\mathrm{Res}\{F(Z_n)h(-iZ_n)\}, \qquad Z_n=\{\pm E_1,\pm E_2\}\\
&=\frac{F(E_{1})-F(-E_{1})}{2E_{1}(E_{1}+E_{2})(E_{1}-E_{2})}-\frac{F(E_{2})-F(-E_{2})}{2E_{2}(E_{1}+E_{2})(E_{1}-E_{2})}\\
&=\frac{1}{2E_{1}E_{2}(E_{1}+E_{2})} \qquad \mathrm{at}\ T=0,
\end{split}
\label{eqn:Matsubara}
\end{equation}
where $\mathrm{Res}\{...\}$ means the residue of the function. So, the expectation value of the loop current at $T=0$ is given by the compact formula
\begin{equation}
\overline{I_{\mathrm{lc}}}=-t'_{\mathrm{nnn}}\frac{e}{\hbar}\frac{S_0}{3}\int \frac{d^2k}{(2\pi)^2} \frac{\mu\,\Xi(\bm{k})\,f_3(\bm{k})}
{E_{1}(\bm{k}) \, E_{2}(\bm{k}) \, [E_{1}(\bm{k})+E_{2}(\bm{k})]}
\label{eqn:I_lc-SC}
\end{equation}
Substituting Eq.~(\ref{eqn:polarization}) into Eq.~(\ref{eqn:I_lc-SC}), the expectation value of the loop current is
\begin{equation}
\overline{I_{\mathrm{lc}}}=-t'_{\mathrm{nnn}}\frac{e}{\hbar}
\frac{2\sqrt{3}}{3} S_0
\int \frac{d^2k}{(2\pi)^2} \frac{\mu\,|\Delta|^2 f^2_3(\bm{k}) }
{E_{1}(\bm{k})\,E_{2}(\bm{k})\,[E_{1}(\bm{k})+E_{2}(\bm{k})]}.
\label{eqn:loop_current_SC}
\end{equation}
The expression under the integral is positive, so the loop current $\overline{I_{\mathrm{lc}}}\neq0$ is clearly nonzero in the presence of the time-reversal-breaking superconducting pairing.

\subsection{The optical Hall spectral weight}

The magneto-optical sum rule (\ref{eqn:sum_rule}) is still applicable in the superconducting state, so the optical Hall spectral weight $W_{\mathrm{Hall}}$ can be related to the commutator of the current operator $\hat{J}_{\alpha}$.  The current operator is defined as a derivative of the BdG Hamiltonian (\ref{eqn:BdG}) with respect to the vector potential:
\begin{equation}
    \hat{J}_{\alpha}=-c\sum_{\bm{k}}\left.\frac{d\hat{H}(\bm{k},\bm A)}{dA_\alpha}\right|_{\bm{A}=0}
= \frac{e}{\hbar}\sum_{\bm{k}}\hat\Psi_{\bm{k}}^{\dagger} \check{V}_{\alpha}(\bm{k})  \hat{\Psi}^{}_{\bm{k}}.
\label{eqn:J_operator}
\end{equation}
Note that the vector potential does not enter the superconducting pairing term $\Delta^{}_{ab}(\bm{k})$ in Eq.~(\ref{eqn:BdG}). Therefore, $\hat J_\alpha \neq d\hat{H}/dk_{\alpha}$ in the superconducting state, unlike in the non-superconducting Haldane's model.  (See a more detailed discussion in Appendix A of Ref.~\refcite{Yakovenko_2009}).  
The $4\times4$ matrix $\check{V}_{\alpha}(\bm{k})$ in Eq.~(\ref{eqn:J_operator}) has the form
\begin{equation}
\check{V}_\alpha(\bm{k})= \begin{pmatrix} \ v_{\alpha}(\bm{k}) & 0 \  \\  \ 0 & v_{\alpha}(\bm{k})\ \end{pmatrix}=\tau_0 \otimes v_{\alpha}(\bm{k}),
\end{equation}
where the $2\times2$ matrix of the velocity $v_\alpha$ is the same as in the normal state,
\begin{equation}
v_\alpha(\bm{k})=\frac {dH_{0}(\bm{k})}{dk_\alpha}
= \begin{pmatrix} \ 0 & u_{\alpha}(\bm{k}) \  
\\  \ u_{\alpha}^*(\bm{k}) & 0 \ \end{pmatrix}
,
\label{dH0/dk}
\end{equation}
and has a form similar to Eq.~(\ref{eqn:velocity_matrix}) with $t_{\mathrm{nnn}}=0$. Thus, the commutator of the current operators reads
\begin{align}
[\hat{J}_{\alpha},\hat{J}_{\beta}]=&\frac{e^2}{\hbar^2}\sum_{\bm{k}}\hat\Psi_{\bm{k}}^{\dagger}[\check{V}_\alpha(\bm{k}),\check{V}_\beta(\bm{k})]\hat{\Psi}^{}_{\bm{k}} \notag \\
=&\frac{e^2}{\hbar^2}\sum_{\bm{k}}\hat\Psi_{\bm{k}}^{\dagger}\tau_0\otimes[v_\alpha(\bm{k}),v_\beta(\bm{k})]\hat{\Psi}^{}_{\bm{k}} 
\label{eqn:current_commutator}
\end{align}
Since we have dropped $t''_{\mathrm{nnn}}$ from $\bm{w}$ in Eq.~(\ref{eqn:w_derivative}), the velocity commutator (\ref{eqn:commutator}) has only the term with $\sigma_3$.  Thus, let us introduce the following (real) function 
\begin{equation}
\Upsilon(\bm k) = \frac{1}{2i} \mathrm{Tr}_\sigma \left\{\sigma_3\, 
\epsilon_{\alpha\beta}\, v_\alpha(\bm k)\, v_\beta(\bm k) \right\}
= - i [u_x(\bm k) \, u_y^*(\bm k) - u_y(\bm k) \, u_x^*(\bm k)],
\label{eqn:Upsilon}
\end{equation}
so that the velocity commutator is
\begin{equation}
[v_\alpha(\bm{k}),v_\beta(\bm{k})]=i \, \sigma_3 \, \epsilon_{\alpha\beta} \, \Upsilon(\bm k).
\label{eqn:[v,v]=Upsilon}
\end{equation}
Using the last line of Eq.~(\ref{eqn:w_cross}), we evaluate $\Upsilon(\bm k)$ as
\begin{equation}
\Upsilon(\bm k) = - \frac{2}{3}S_0t_{\rm{nn}}^2 \, f_3(\bm{k}).
\label{eqn:Upsilon-f3}
\end{equation}
The velocity commutator contains the same function $f_3(\bm{k})$ as shown in Fig.~\ref{fig:Berry}.

Substituting Eq.~(\ref{eqn:[v,v]=Upsilon}) into Eq.~(\ref{eqn:current_commutator}), the commutator of the current operators is
\begin{equation}
\frac{[\hat{J}_{\alpha},\hat{J}_{\beta}]}{S} = i \epsilon_{\alpha\beta} 
\frac{e^2}{\hbar^2} \int\frac{d^2k}{(2\pi)^2} \, \Upsilon(\bm k) \, 
\hat\Psi_{\bm{k}}^{\dagger}\tau_0\otimes\sigma_3\hat{\Psi}^{}_{\bm{k}}
\label{eqn:JJ-Ups}
\end{equation}
in agreement with Eq.~(39) of Ref.~\refcite{Yakovenko}.  Then, the expectation value of Eq.~(\ref{eqn:JJ-Ups}) is
\begin{equation}
\frac{\langle [\hat{J}_{\alpha},\hat{J}_{\beta}]\rangle}{S}
=i\epsilon_{\alpha\beta}\frac{e^2}{\hbar^2}
T\sum_{\nu}\int\frac{d^2k}{(2\pi)^2} \Upsilon(\bm{k}) \,
\mathrm{Tr}_{\tau\sigma}\left\{\tau_0\otimes\sigma_3 \, \check{G}(\nu,\bm{k})\right\}.
\label{eqn:current_commutator2}
\end{equation}
Only the term $G_{0,3}$ in Eq.~(\ref{eqn:All_Green}) with the structure $\tau_0\otimes\sigma_3$ contributes to $\check{G}(\nu,\bm{k})$ in Eq.~(\ref{eqn:current_commutator2}). Therefore, the expectation value in Eq.~(\ref{eqn:current_commutator2}) is similar to Eq.~(\ref{eqn:loop_PRX})
\begin{equation}
\frac{\langle[\hat{J}_{\alpha},\hat{J}_{\beta}]\rangle}{S}
=i\epsilon_{\alpha\beta}\frac{2e^2}{\hbar^2}
T\sum_{\nu}\int\frac{d^2k}{(2\pi)^2}\frac{\mu \, \Xi(\bm{k}) \, \Upsilon(\bm{k})}{[\hbar^2\nu^2+E_{1}^2(\bm{k})]\,[\hbar^2\nu^2+E_{2}^2(\bm{k})]}.
\end{equation}
Evaluating the Matsubara sum as in Eq.~(\ref{eqn:Matsubara}), we find a compact formula
\begin{equation}
\frac{\langle[\hat{J}_{\alpha},\hat{J}_{\beta}]\rangle}{S}
=i\epsilon_{\alpha\beta}\frac{e^2}{\hbar^2}
\int\frac{d^2k}{(2\pi)^2}\frac{\mu \, \Xi(\bm{k}) \, \Upsilon(\bm{k})}{E_{1}(\bm{k})\,E_{2}(\bm{k}) \, [E_{1}(\bm{k})+E_{2}(\bm{k})]}.
\label{eqn:currents-Ups}
\end{equation}
Substituting $\Xi(\bm{k})$ from Eq.~(\ref{eqn:polarization}) and $\Upsilon(\bm{k})$ from Eq.~(\ref{eqn:Upsilon-f3}) into Eq.~(\ref{eqn:currents-Ups}), we obtain the final formula for our model
\begin{equation}
\frac{\langle[\hat{J}_{\alpha},\hat{J}_{\beta}]\rangle}{S}
=-i\epsilon_{\alpha\beta}\frac{e^2}{\hbar^2}\frac{4\sqrt{3}}{3} S_0t_{\rm{nn}}^2
\int\frac{d^2k}{(2\pi)^2} \frac{\mu\,|\Delta|^2 \, f_3^2(\bm{k})}
{E_{1}(\bm{k})\,E_{2}(\bm{k})\,[E_{1}(\bm{k})+E_{2}(\bm{k})]}.
\label{eqn:currents_SC}
\end{equation}
Comparing Eqs.~(\ref{eqn:loop_current_SC}) and (\ref{eqn:currents_SC}), we relate the loop current $\overline{I_{\mathrm{lc}}}$ with the current commutator and the optical Hall spectral weight $W_{\mathrm{Hall}}$ via Eq.~(\ref{eqn:sum_rule}):
\begin{equation}
\overline{I_{\mathrm{lc}}}
=-\frac{i}{4}\frac{t'_{\mathrm{nnn}}}{t^2_{\mathrm{nn}}}\frac{\hbar}{e}\frac{\langle [\hat{J}_{\alpha},\hat{J}_{\beta}]\rangle}{S}\epsilon_{\alpha\beta}
= 2\frac{t'_{\mathrm{nnn}}}{t^2_{\mathrm{nn}}}\frac{1}{e}\, W_{\mathrm{Hall}}.
\label{eqn:loop_Hall}
\end{equation}
Equation (\ref{eqn:loop_Hall}) for the superconducting model is the same as Eq. (\ref{eqn:currents_full}) for Haldane's model.  Moreover, comparing Eqs.~(\ref{eqn:JJ-Ups}) and (\ref{eqn:Upsilon-f3}) with Eqs.~(\ref{eqn:SC_lc_operator}) and (\ref{eqn:loop_tensor}), we observe that the first equality in Eq.~(\ref{eqn:loop_Hall}) is satisfied not only for the expectation values, but also for the second-quantized operators themselves
\begin{equation}
\hat{I}_{\mathrm{lc}}
=-\frac{i}{4}\frac{t'_{\mathrm{nnn}}}{t^2_{\mathrm{nn}}}\frac{\hbar}{e}
\frac{[\hat{J}_{\alpha},\hat{J}_{\beta}]}{S}\epsilon_{\alpha\beta}.
\label{eqn:loop-commut-operators}
\end{equation}
We see that the products of the two nearest-neighbor hopping operators in the commutator on the right-hand side of Eq.~(\ref{eqn:loop-commut-operators}) are equivalent to the next-nearest-neighbor hopping operators on the left-hand side.

\subsection{The ac Hall conductivity}

The antisymmetric tensor of the intrinsic Hall conductivity can be found explicitly from Eq.~(\ref{eqn:J_operator}) using the Feynman diagram in Fig.~\ref{fig:Feynman}:
\begin{equation}
\sigma_{H}(\omega)=i\frac{\epsilon_{\alpha\beta}}{2}\frac{e^2}{\hbar^3\omega}
\lim_{i\omega_n\to\omega+i0^+} T\sum_\nu\int \frac{d^2k}{(2\pi)^2}
\mathrm{Tr}_{\tau\sigma}\{\check{V}_{\alpha}\check{G}(\nu)\check{V}_{\beta}\check{G}(\nu+\omega_n)\}.
\label{eqn:Hall_definition_SC}
\end{equation}
Expanding Green's function in terms of the Pauli matrices $\tau_q$ in the Nambu space as $\check{G}(\nu)=\sum_{q=0}^3\tau_q\otimes G_q(\nu)$ and then taking the trace in the Nambu space, we get the following
\begin{equation}
\sigma_{H}(\omega)=i\epsilon_{\alpha\beta}\frac{e^2T}{\hbar^3\omega}
\lim_{i\omega_n\to\omega+i0^+} \sum_\nu\int \frac{d^2k}{(2\pi)^2}
\mathrm{Tr}_\sigma \sum_{q=0}^3 v_\alpha(\bm{k}) G_q(\nu,\bm{k})v_\beta(\bm{k}) G_q(\nu+\omega_n,\bm{k}).
 \label{eqn:Hall_Gj}
\end{equation}
Taking the trace in the sublattice space and evaluating the Matsubara sum as described in Appendix \ref{App:Matsubara}, we obtain the following expression

\begin{equation}
\sigma_{H}(\omega)=\frac{e^2}{\hbar\omega}\lim_{i\omega_n\to \omega+i0^+}\int \frac{d^2k}{(2\pi)^2}\, \Upsilon(\bm{k}) \, M(\omega_n,\bm{k}) \, \Xi(\bm{k}) \, \mu,
\label{eqn:Hall}
\end{equation}
where the function $M(\omega_n)$ at $T=0$ is
\begin{equation}
M=\frac{i\omega_n}{E_1E_2(E_2+E_1)[\omega_n^2+(E_2+E_1)^2]}.
\label{eqn:M-final}
\end{equation}
Carrying out the analytic continuation in Eqs.~(\ref{eqn:Hall}) and (\ref{eqn:M-final}), we find the real and imaginary parts of ac Hall conductivity at zero temperature
\begin{align}
\mathrm{Re}\,\sigma_{H}(\omega)=&\frac{e^2}{h}\frac{2}{\pi}\int d^2 k\,  \frac{L_{H}(\bm{k})}{[E_2(\bm k)+E_1(\bm k)]^2-\omega^2}, \label{eqn:Real_Hall} \\
\mathrm{Im}\,\sigma_{H}(\omega)=\frac{e^2}{h}&\frac{1}{\omega}\int d^2 k\,
  L_{H}(\bm{k}) \, \delta\{\omega-[E_2(\bm k)+E_1(\bm k)]\}.
\label{eqn:Imag_Hall}
\end{align}
Here we introduced the Hall spectral density $L_{H}(\bm{k})$ for our superconducting model
\begin{equation}
L_{H}(\bm{k})=\frac{Q_H(\bm{k})}
{4\,E_{1}(\bm{k})\,E_{2}(\bm{k})\,[E_{1}(\bm{k})+E_{2}(\bm{k})]},
\label{eqn:L}
\end{equation}
where the numerator
\begin{equation}
Q_H(\bm{k})= \mu \, \Upsilon(\bm{k}) \, \Xi(\bm{k})
= \frac{i}{2} \mathrm{Tr}_\sigma \left\{
\epsilon_{\alpha\beta}\, v_\alpha(\bm k)\, v_\beta(\bm k) \,
[\Delta(\bm{k}),\Delta^\dagger(\bm{k})] \, H_{0}(\bm{k})
\right\}
\label{eqn:Q}
\end{equation}
can be expressed as a trace of a product of the velocity commutator and the commutator of the superconducting pairing potentials.  Equations (\ref{eqn:Real_Hall}), (\ref{eqn:Imag_Hall}) and (\ref{eqn:L}) are similar to Eqs.~(7) and (8) obtained in Ref.~\refcite{Kallin}, while the numerator (\ref{eqn:Q}) is reminiscent of the numerator in Eq.~(36) of Ref.~\refcite{Yakovenko}.  (A more general case was studied in Refs.~\refcite{Kallin_2017,Brydon} to include spin-orbit interaction, which is omitted in our model.)

Substituting Eq.~(\ref{eqn:Imag_Hall}) into Eq.~(\ref{eqn:weight}), we find that the optical Hall spectral weight $W_{\mathrm{Hall}}$ is an integral of the Hall spectral density $L_{H}(\bm{k})$ over the Brillouin zone:
\begin{equation}
W_{\mathrm{Hall}}=\frac{e^2}{\hbar} \int \frac{d^2k}{(2\pi)^2}  L_{H}(\bm{k}).
\label{eqn:Weight_SC}
\end{equation}
Comparing Eq.~(\ref{eqn:Weight_SC}) with Eq.~(\ref{eqn:weight_full}), we see that the function $L_{H}(\bm{k})$ for the chiral superconductor plays a role similar to the current kernel $K_H(\bm{k})/2w(\bm k)$ for Haldane's model. Comparing Eq.~(\ref{eqn:L}), (\ref{eqn:Q}) and (\ref{eqn:Weight_SC}) with Eq.~(\ref{eqn:currents-Ups}), we confirm the magneto-optical sum rule (\ref{eqn:sum_rule}).

 Substituting $\Xi(\bm{k})$ from Eq.~(\ref{eqn:polarization}) and $\Upsilon(\bm{k})$ from Eq.~(\ref{eqn:Upsilon-f3}) into Eq.~(\ref{eqn:Q}), we find that the numerator $Q_H(\bm k)$ is proportional to $f_3^2(\bm{k})$ in our model
\begin{equation}
Q_H(\bm{k})= \mu \, \Upsilon(\bm{k}) \, \Xi(\bm{k})
= - \frac{4\sqrt{3}}{3} \, S_0\,t_{\rm{nn}}^2 \, \mu \, |\Delta|^2 f_3^2(\bm{k}).
\label{eqn:Q-f3}
\end{equation}

\subsection{Optical peak}

\begin{figure}[!t]
\centering
\begin{subfigure}[t]{0.45\textwidth}
    \includegraphics[width=\textwidth]{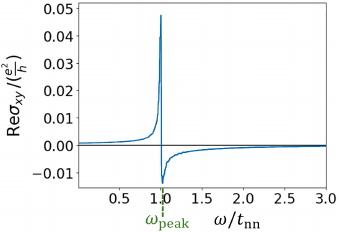}
    \caption{}
\end{subfigure}
~
\begin{subfigure}[t]{0.45\textwidth}
    \includegraphics[width=\textwidth]{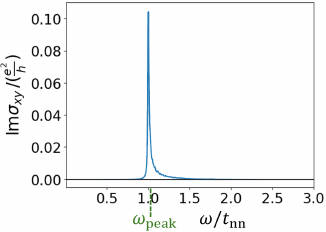}
    \caption{}
\end{subfigure}
\caption{Frequency-dependence of the Hall conductivity in clean chiral d-wave superconductor with 2D honeycomb lattice, with $\mu=-0.5t_{\mathrm{nn}}$ and $\Delta=0.1t_{\mathrm{nn}}$. (a) is the imaginary part and (b) is the real part.}
\label{fig:peak}
\end{figure}

The frequency dependencies of the real and imaginary parts of $\sigma_{H}(\omega)$ calculated from Eqs.~(\ref{eqn:Real_Hall}) and (\ref{eqn:Imag_Hall}) are shown in Fig.~\ref{fig:peak} and are consistent with Fig.~6 in Ref.~\refcite{Yakovenko}.  We observe a sharp peak  in the imaginary (absorptive) part $\mathrm{Im}\,\sigma_{H}(\omega)$ at the frequency $\omega_{\mathrm{peak}}\approx2\mu$.  On the other hand, there are no features at the frequency $\omega=2\Delta$ corresponding to transitions across the superconducting gap.

\begin{figure}[!t]
\centering
\includegraphics[width=0.6\textwidth]{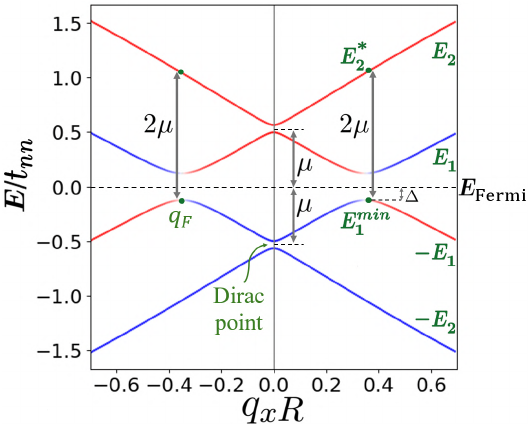}
\caption{Vertical axis: Energy eigenvalues of the BdG Hamiltonian near the $K$ point of the honeycomb lattice.  Horizontal axis: The deviation $q_x=k_x-K_x$ of momentum from the $K$ point in the $x$ direction.  The horizontal dashed line represents the Fermi level at the energy distance $\mu$ from the Dirac point.  Blue and red colors represent electron and hole states, respectively.  The vertical lines with arrows indicate optical transitions from the energy $-E_1$ in the lower band to $E_2$ in the upper band.}
\label{fig:Band}
\end{figure}

To explain the origin of the sharp peak in Fig.~(\ref{fig:peak}), let us examine the BdG energy spectrum shown in Fig.~\ref{fig:Band} in the vicinity of the $K$ point in the Brillouin zone.  The dashed horizontal line in Fig.~\ref{fig:Band} corresponds to the Fermi level, which is taken as zero energy in the BdG formalism.  The delta function in Eq.~(\ref{eqn:Imag_Hall}) only allows interband electron excitations from the occupied band at the energy $-E_1$ to the empty band at $E_2$ as shown in Fig.~\ref{fig:Band}, or from $-E_2$ to $E_1$ (not shown).  Consistent with the symmetry constraints discussed in Ref.~\refcite{Ahn2021}, intraband excitations from $-E_1$ to $E_1$ and from $-E_2$ to $E_2$ are forbidden.  (Intraband transitions become permitted when spin-orbit interaction is included, as discussed in Ref.~\refcite{Ahn2021} and found in Eq.~(31) of Ref.~\refcite{Kallin_2017} and in Fig.~2 of Ref.~\refcite{Brydon}.)  The maximum of $\mathrm{Im}\,\sigma_{H}(\omega)$ in Eq.~(\ref{eqn:Imag_Hall}) is obtained by maximizing $L_{H}(\bm{k})$ under the integral.  The maximum of $L_{H}(\bm{k})$ is achieved when $E_1(\bm{k})$ in the denominator of Eq.~(\ref{eqn:L}) is minimized, assuming by convention that $E_1(\bm{k})<E_2(\bm{k})$.  The BdG energy $E_1(\bm{k})$ has its minimal value $E_1^{\mathrm{min}}(\bm{k})=\Delta$ on the normal-state Fermi surface, labeled as $q_F$ in Fig.~\ref{fig:Band}.  Here we introduced the momentum $\bm{q}=\bm{k}-\bm{K}$ measured from the Dirac point $\bm{K}$, so that $q_F$ is the normal-state Fermi momentum.  The corresponding energy of the excited date for the ``vertical'' optical transitions is $E_2^*=2\mu$, as shown in Fig.~\ref{fig:Band}.  Thus, in agreement with Fig.~\ref{fig:peak}(b), the peak frequency is $\omega_{\mathrm{peak}}=E^{min}_1+E_2^*\approx\Delta+2\mu\approx2\mu$, if $\Delta\ll2\mu$.  

Alternatively, this result can be understood as follows.  Time-reversal symmetry breaking originates from the superconducting pairing $\Delta$, which modifies electronic properties only in the narrow vicinity of the normal-state Fermi surface.  Thus, optical transitions in the time-reversal-odd Hall channel are only possible from the normal-state Fermi surface at the energy $-\mu$ below the Dirac point to the upper band at the energy $+\mu$ above the Dirac point, thus corresponding to the optical frequency $\omega_{\mathrm{peak}}\approx2\mu$.

\begin{figure}[!b]
\centering
\hfill
\begin{subfigure}[t]{0.45\textwidth}
    \includegraphics[width=\textwidth]{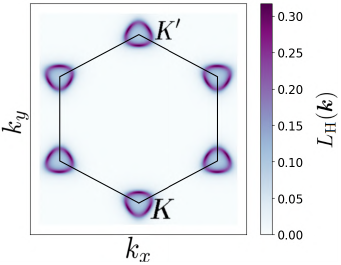}
    \caption{}
\end{subfigure}
\hfill
\begin{subfigure}[t]{0.35\textwidth}
    \includegraphics[width=\textwidth]{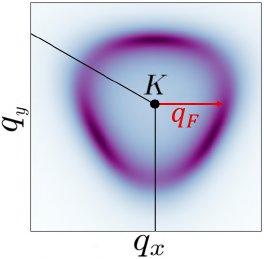}
    \caption{}
\end{subfigure}
\hfill{}
\caption{(a) Spectral density $L_{H}(\bm{k})$ from Eq.~(\ref{eqn:L}) shown over the Brillouin zone. (b) A close-up in the vicinity of the $K$ point.}
\label{fig:FS}
\end{figure}

Figure \ref{fig:FS} illustrates that, indeed, the function $L_{H}(\bm{k})$ in Eq.~(\ref{eqn:L}) is maximal on the normal-state Fermi surface in the 2D Brillouin zone.  The Fermi surface is approximately circular when $q_FR\ll1$.  Near the Fermi surface, we can write $E_1\approx\sqrt{\Delta^2+(v_F\delta q)^2}$ and $E_2=2\mu+v_F\delta q$, where $v_F$ is the Dirac velocity, and $\delta q=q-q_F$ is the momentum deviation from $q_F$ in the direction perpendicular to the normal-state Fermi surface. Substituting into Eqs.~(\ref{eqn:Imag_Hall}), (\ref{eqn:L}), and (\ref{eqn:Weight_SC}), it can be shown that the height of the peak, $\mathrm{Im}\,\sigma_{H}(\omega_{\mathrm{peak}})\sim\Delta$, and its frequency width, $\delta\omega\sim\Delta$, in Fig.~\ref{fig:peak}(b) are both proportional to $\Delta$, while the spectral weight, $W_{\mathrm{Hall}}\sim\Delta^2$, is proportional to $\Delta^2$, assuming $\Delta\ll2\mu$.

The sharpness of the peak is a consequence of the fact that the sum of the lower and upper energies $E_1(\bm k)+E_2(\bm k)$ is constant for all $\bm k$ on the normal-state Fermi surface in the 2D Brillouin zone.  Thus, the integration over momentum in Eq.~(\ref{eqn:Imag_Hall}) results in a strong enhancement of the output in the narrow frequency range $\delta\omega\sim\Delta$ at $\omega_{\mathrm{peak}}$.  The above property of the constant $E_1(\bm k)+E_2(\bm k)$ on the normal-state Fermi surface follows from the particle-hole symmetry of the normal-state energy spectrum for the honeycomb lattice.  For a small $q_F$, this also follows from the approximate circular symmetry of the Fermi surface.  

The real and imaginary parts of $\sigma_{H}(\omega)$ in Eqs.~(\ref{eqn:Real})-(\ref{eqn:Imag}) and (\ref{eqn:Real_Hall})-(\ref{eqn:Imag_Hall}) satisfy the Kramers-Kronig relation
\begin{equation}
\mathrm{Re}\,\sigma_{H}(\omega)=\frac{2}{\pi}\int_0^\infty d\omega' \, \frac{\omega'\,\mathrm{Im}\,\sigma_{H}(\omega')}{\omega'^2-\omega^2}.
\label{eqn:KK}
\end{equation}
Given that $\mathrm{Im}\,\sigma(\omega)$ has a narrow peak at $\omega=\omega_{\mathrm{peak}}$ in Fig.~\ref{fig:peak}(b), the variable $\omega'$ under the integral in Eq.~(\ref{eqn:KK}) can be approximately replaced by $\omega_{\mathrm{peak}}$, so that
\begin{equation}
\mathrm{Re}\,\sigma_{H}(\omega)\approx \frac{2}{\pi}\frac{1}{\omega_{\mathrm{peak}}^2-\omega^2}\int_0^\infty d\omega' \, \omega'\,\mathrm{Im}\,\sigma_{H}(\omega')=\frac{4\,W_{\mathrm{Hall}}}{\omega_{\mathrm{peak}}^2-\omega^2}.
\label{eqn:Real_approx}
\end{equation}
Equation (\ref{eqn:Real_approx}) gives an appropriate formula for $\mathrm{Re}\,\sigma_{H}(\omega)$, which is consistent with Fig.~\ref{fig:peak}(a) and is applicable for $\omega$ not too close to $\omega_{\mathrm{peak}}$.  Using Eq.~(\ref{eqn:Real_approx}), we find the high- and low frequency limits for $\mathrm{Re}\,\sigma_{H}(\omega)$:
\begin{align}
\sigma_{H}(0) &\approx  \frac{4\,W_{\mathrm{Hall}}}{\omega_{\mathrm{peak}}^2},
\label{eqn:low} \\
\mathrm{Re}\,\sigma_{H}(\omega\gg\omega_{\mathrm{peak}})
&\approx - \frac{4\,W_{\mathrm{Hall}}}{\omega^2}.
\label{eqn:high}
\end{align}
The dc Hall conductivity $\sigma_{H}(0)$ is necessarily real.  The high-frequency  limit (\ref{eqn:high}) was derived in Ref.~\refcite{Shastry_1993} in terms of the expectation value of the current commutator, which is related to $W_{\mathrm{Hall}}$ via the magneto-optical sum rule\cite{Kotliar} in Eq.~(\ref{eqn:sum_rule}).  Comparing Eqs.~(\ref{eqn:low}) and (\ref{eqn:high}), we find a general relation between the high- and low frequency limits for $\mathrm{Re}\,\sigma_{H}(\omega)$:
\begin{equation}
\mathrm{Re}\,\sigma_{H}(\omega\gg\omega_{\mathrm{peak}})
\approx -\frac{\omega_{\mathrm{peak}}^2}{\omega^2}\,\sigma_{H}(0).
\label{eqn:high-low}
\end{equation}
A similar relation was recently discussed in Refs.~\refcite{Armitage_2025,Armitage_2026}.

Direct electric measurement of dc Hall conductivity $\sigma_{H}(0)$ is not possible in superconductors due to shunting by supercurrent.  However, an indirect indicator of the dc AHE in a chiral superconductor was recently proposed in Ref.~\refcite{Levchenko_2026} using the Coulomb drag effect in an adjacent normal layer.

Fig.~\ref{fig:Band} also shows a gap opening at the Dirac point $K$ (and $K'$) away from the normal-state Fermi energy, which was discussed in detail in Ref.~\refcite{Yakovenko}.  The origin of this Dirac gap can be qualitatively understood as follows.  As shown in Fig.~\ref{fig:Pairing}, one of the superconducting pairing amplitudes $\Delta_{BA}(\bm{k})$ and $\Delta_{AB}(\bm{k})$ vanishes at the point $K$ and another at the point $K'$.  Thus, treating $\Delta$ in the second-order perturbation theory results in a non-zero energy shift for sublattice $A$, but no shift for sublattice $B$, or vice versa.  As a result, the energies of sublattices $A$ and $B$ split at the Dirac points, which is equivalent to opening of an energy gap, as found in Refs.~\refcite{Yakovenko} and \refcite{Sigrist}.

\section{Conclusions}

For Haldane's model, we found that a non-zero expectation value of loop currents in Eq.~(\ref{eqn:full_loop_current}) requires the presence of both real $t_{\mathrm{nnn}}'$ and imaginary $t_{\mathrm{nnn}}''$ parts of the next-nearest-neighbor tunneling amplitudes.  In contrast, only the imaginary part $t_{\mathrm{nnn}}''$ is needed for a non-zero dc quantum Hall effect.  Thus, the requirements for loop currents and the quantum Hall effect are somewhat different.

We found a direct relation (\ref{eqn:currents_full}) between the expectation value of the loop currents $\overline{I_{\mathrm{lc}}}$ and the optical Hall spectral weight $W_{\mathrm{Hall}}$ defined in Eq.~(\ref{eqn:weight}).  Using this relation, the presence and magnitude of stationary loop currents can be deduced from measurements of the optical Hall conductivity.

We also introduced the concept of an antisymmetric velocity kernel in Eq.~(\ref{eqn:K_H}).  It determines both the ac and dc Hall conductivities in Eqs.~(\ref{eqn:Real}) and (\ref{eqn:Imag}) and the optical Hall spectral weight in Eq.~(\ref{eqn:weight_full}).  It is similar but more general than the Berry curvature, which applies only to the dc case in Eq.~(\ref{eqn:DC_Hall}).

For a chiral $d_{x^2-y^2}\pm id_{xy}$ superconductor on the honeycomb lattice, we clarified and extended the results of Ref.~\refcite{Yakovenko}.  The expectation value of loop currents is expressed as an integral over the Brilloun zone in Eq.~(\ref{eqn:loop_current_SC}) and is related to the Hall spectral weight in Eq.~(\ref{eqn:loop_Hall}).  The real and imaginary parts of ac Hall conductivity (\ref{eqn:Real_Hall}) and (\ref{eqn:Imag_Hall}) and the Hall spectral weight (\ref{eqn:Weight_SC}) are expressed in terms of the Hall spectral density $L_{H}(\bm{k})$ introduced in Eq.~(\ref{eqn:L}).  

The calculated frequency dependence of the optical Hall conductivity demonstrates a sharp peak at $\omega_{\mathrm{peak}}\approx2\mu$ in Fig.~\ref{fig:peak}.  The peak is due to interband transitions from the electronic states corresponding to the normal-state Fermi surface at the energy $-\mu$ below the Dirac point to the upper band at the energy $+\mu$ above the Dirac point.  These transitions in the Hall channel are activated by the time-reversal-breaking superconducting pairing.  Figure \ref{fig:peak} indicates that the optimal frequency for the observation of the ac Hall conductivity is $\omega_{\mathrm{peak}}$.  In contrast, most previous measurements\cite{Fried_2014,Paglione_2021} of the ac Hall signal were made at much higher frequencies $\omega\gg\omega_{\mathrm{peak}}$, where it is strongly suppressed.  Recent measurements\cite{Armitage_2025,Blumberg_2024} at lower frequencies may be closer to the maximum signal.

In conclusion, measurements of the optical Hall conductivity can provide useful information about the presence and magnitude of steady loop currents for both Haldane's model and time-reversal-breaking superconductors.

After this paper was posted on arXiv, its AI summary was automatically generated at \url{https://gist.science/paper/2608.22083} by Gist.Science.  Produced without any request or participation of the authors, it represents a good and useful popular summary of the paper for the general audience.


\section*{ORCID}

\noindent Azzam Alzahrani - \url{https://orcid.org/0009-0008-1017-3980}

\noindent Victor M. Yakovenko - \url{https://orcid.org/0000-0003-3754-1794}

\appendix{Velocity kernel}
\label{App:velocity}

Using Eq.~(\ref{eqn:w_derivative}), the cross product in Eq.(\ref{eqn:K_function}) can be evaluated as
\begin{equation}
\begin{split}
\left(\frac{d\bm{w}}{dk_{x}}\times\frac{d\bm{w}}{dk_{y}}\right)_1=&2t^{}_{\mathrm{nn}}t''_{\mathrm{nnn}}\left(\frac{df_2}{dk_x}\frac{df_3}{dk_y}-\frac{df_3}{dk_x}\frac{df_2}{dk_y}\right)
\\ =&2t^{}_{\mathrm{nn}}t''_{\mathrm{nnn}}\sum_{j,j'} [R_{j,x}c_{j',y}\cos(\bm{k}\cdot \bm{R}_{j})\cos(\bm{k}\cdot \bm{c}_{j'})-c_{j,x}R_{j',y} \cos(\bm{k}\cdot \bm{c}_{j})\cos(\bm{k}\cdot \bm{R}_{j'}) ]
\\ =&2t^{}_{\mathrm{nn}}t''_{\mathrm{nnn}}\sum_{j,j'} (R_{j,x}c_{j',y}-R_{j,y}c_{j',x})\cos(\bm{k}\cdot \bm{R}_{j})\cos(\bm{k}\cdot \bm{c}_{j'})
\\=& 2t^{}_{\mathrm{nn}}t''_{\mathrm{nnn}}\sum_{j,j'} Rc\left[\cos(\phi_j)\sin(\theta_{j'})-\sin(\phi_j)\cos(\theta_{j'})\right]\cos(\bm{k}\cdot \bm{R}_{j})\cos(\bm{k}\cdot \bm{c}_{j'}) 
\\=& 2t^{}_{\mathrm{nn}}t''_{\mathrm{nnn}}\sum_{j,j'} Rc\sin(\theta_{j'}-\phi_{j})\cos(\bm{k}\cdot \bm{R}_{j})\cos(\bm{k}\cdot \bm{c}_{j'}),
\\ \left(\frac{d\bm{w}}{dk_{x}}\times\frac{d\bm{w}}{dk_{y}}\right)_2=&2t^{}_{\mathrm{nn}}t''_{\mathrm{nnn}}\left(\frac{df_3}{dk_x}\frac{df_1}{dk_y}-\frac{df_1}{dk_x}\frac{df_3}{dk_y}\right)
\\ =&2t^{}_{\mathrm{nn}}t''_{\mathrm{nnn}}\sum_{j,j'} -[c_{j,x}R_{j',y}\cos(\bm{k}\cdot \bm{c}_{j})\sin(\bm{k}\cdot \bm{R}_{j'})-R_{j,x}c_{j',y} \sin(\bm{k}\cdot \bm{R}_{j})\cos(\bm{k}\cdot \bm{c}_{j'}) ]
\\ =&2t^{}_{\mathrm{nn}}t''_{\mathrm{nnn}}\sum_{j,j'} (R_{j,x}c_{j',y}-R_{j,y}c_{j',x})\sin(\bm{k}\cdot \bm{R}_{j})\cos(\bm{k}\cdot \bm{c}_{j'})
\\=& 2t^{}_{\mathrm{nn}}t''_{\mathrm{nnn}}\sum_{j,j'} Rc\left[\cos(\phi_j)\sin(\theta_{j'})-\sin(\phi_j)\cos(\theta_{j'})\right]\sin(\bm{k}\cdot \bm{R}_{j})\cos(\bm{k}\cdot \bm{c}_{j'})
\\=& 2t^{}_{\mathrm{nn}}t''_{\mathrm{nnn}}\sum_{j,j'} Rc\sin(\theta_{j'}-\phi_{j})\sin(\bm{k}\cdot \bm{R}_{j})\cos(\bm{k}\cdot \bm{c}_{j'}),
\\\left(\frac{d\bm{w}}{dk_{x}}\times\frac{d\bm{w}}{dk_{y}}\right)_3=&2t^{}_{\mathrm{nn}}t''_{\mathrm{nnn}}\left(\frac{df_1}{dk_x}\frac{df_2}{dk_y}-\frac{df_2}{dk_x}\frac{df_1}{dk_y}\right)
\\=&t_{\mathrm{nn}}^2\sum_{j,j'} R_{j,x}R_{j',y}[\sin(\bm{k}\cdot \bm{R}_{j'})\cos(\bm{k}\cdot \bm{R}_{j})- \sin(\bm{k}\cdot \bm{R}_j) \cos(\bm{k}\cdot \bm{R}_{j'})].
\\ =&t_{\mathrm{nn}}^2\sum_{j<j'}R^2\sin(\phi_{j'}-\phi_{j})\sin[\bm{k}\cdot(\bm{R_{j'}}-\bm{R_{j}})]
\\ =&-t_{\mathrm{nn}}^2 R^2 \frac{\sqrt{3}}{2} \sum_{j}\sin(\bm{k}\cdot \bm{c}_j)
=-t_{\mathrm{nn}}^2 R^2 \frac{\sqrt{3}}{2}\,f_3(\bm k)
=-t_{\mathrm{nn}}^2 \frac{S_0}{3}\,f_3(\bm k),
\end{split}
\label{eqn:w_cross}
\end{equation}
where $\phi_{j}\in\{0,\frac{2\pi}{3},\frac{4\pi}{3}\}$ and $\theta_{j}\in\{\frac{3\pi}{6},\frac{7\pi}{6},\frac{11\pi}{6}\}$ are the angles that $\bm{R}_j$ and $\bm{c}_j$, respectively, make with the $+x$-axis.  We also expressed the last line in Eq.~(\ref{eqn:w_cross}) in terms of the unit cell area $S_0$ given by Eq.~(\ref{eqn:unit_cell}).

Multiplying Eq.~(\ref{eqn:w_cross}) by the vector $\bm{w}$ from Eq.~(\ref{eqn:W}), we find
\begin{equation}
\begin{split}
K_{\alpha\beta}(\bm{k})=&\bm{w}(\bm{k})\cdot\left[\frac{\partial\bm{w}(\bm{k})}{\partial k_{\alpha}}\times\frac{\partial\bm{w}(\bm{k})}{\partial k_{\beta}}\right]=Rct^2_{\mathrm{nn}}t''_{\mathrm{nnn}} \epsilon_{\alpha\beta}  \\
\times&\left\{-f^2_3+\sum_{i,j,j'}\sin(\theta_{j'}-\phi_j)\cos(\bm{k}\cdot\bm{c}_{j'})[\sin(\bm{k}\cdot\bm{R}_i)\sin(\bm{k}\cdot\bm{R}_j)+\cos(\bm{k}\cdot\bm{R}_i)\cos(\bm{k}\cdot\bm{R}_j)]\right\}\\
=&\epsilon_{\alpha\beta}Rct^2_{\mathrm{nn}}t''_{\mathrm{nnn}}\left\{\left[\sum_{i}\cos(\bm{k}\cdot\bm{c}_{i})\right]^2-\left[\sum_{i}\sin(\bm{k}\cdot\bm{c}_{i})\right]^2\right\}\\
=&\epsilon_{\alpha\beta}\frac{2}{3}S_0t^2_{\mathrm{nn}}t''_{\mathrm{nnn}} \left[f^2_0(\bm{k})-f_3^2(\bm{k})\right].
\end{split}
\end{equation}

\appendix{BdG Green's functions}
\label{App:Green}

Using the block decomposition of the Hamiltonian in Eq.~(\ref{eqn:BdG_general}), Green's function
\begin{equation}
\check{G}_{}(\nu,\bm{k})=[i\nu \, \check{\mathbb{1}}-\check{H}_{}(\bm{k})]^{-1}.
\end{equation}
can be expressed as
\begin{equation}
    \check{G}_{}(\nu,\bm{k})=\begin{pmatrix} \ [(G^p_{0})^{-1}-\Sigma^h]^{-1}  & [(G^p_{0})^{-1}-\Sigma^h]^{-1}\Delta G^h_{0}\  \\  \ [(G^h_{0})^{-1}-\Sigma^p]^{-1}\Delta^\dagger G^p_{0} & [(G^h_{0})^{-1}-\Sigma^p]^{-1}\ \end{pmatrix}.
\label{eqn:Green_SC}
\end{equation}
The normal-state hole and particle Green's functions,
\begin{equation}
\begin{split}
G^p_{0}=&[i\nu \, \sigma_0-H_{0}]^{-1}, \\
G^h_{0}=&[i\nu \, \sigma_0+H_{0}]^{-1},
\end{split}
\end{equation}
have the same form introduced in Eq.~(\ref{eqn:Green}). We define the self energies as
\begin{equation}
\begin{split}
\Sigma^p=&\Delta^\dagger G^p_{0}\Delta^{}, \\
\Sigma^h=&\Delta^{} G^h_{0}\Delta^\dagger.
\end{split}
\end{equation}
Using $\Delta$ from Eq.~(\ref{eqn:Pairing}) and Green's functions $G^p_{0}$ and $G^h_{0}$ from Eq.~(\ref{eqn:Green}), we get
\begin{equation}
\begin{split}
\Sigma^h=&\begin{pmatrix} \ -|\Delta^{}_{AB}|^2(i\nu+w_0+w_3) & \Delta^{}_{AB}\Delta^{*}_{BA}(w_1-iw_2) \  \\  \ \Delta^{*}_{AB}\Delta^{}_{BA}(w_1+iw_2) & -|\Delta^{}_{BA}|^2(i\nu+w_0-w_3)\ \end{pmatrix}\frac{1}{D_{0,h}}, \\
\Sigma^p=&\begin{pmatrix} \ -|\Delta^{}_{BA}|^2(i\nu-w_0-w_3) & -\Delta^{}_{AB}\Delta^{*}_{BA}(w_1-iw_2) \  \\  \ -\Delta^{*}_{AB}\Delta^{}_{BA}(w_1+iw_2) & -|\Delta^{}_{AB}|^2(i\nu-w_0+w_3)\ \end{pmatrix}\frac{1}{D_{0,p}},
\end{split}
\end{equation}
where $D_{0,p/h}=(i\nu-\varepsilon^{p/h}_1)(i\nu-\varepsilon^{p/h}_2)$. The normal-state eigenenergies of particles and holes are $\varepsilon^{p}_{1/2}=-\mu\pm w$ and $\varepsilon^{h}_{1/2}=\mu\pm w$, respectively.
For the normal-state Hamiltonian in Eq.~(\ref{eqn:Hamiltonian}), we find the following
\begin{equation}
\begin{split}
&[(G^p_{0})^{-1}-\Sigma^h]^{-1}=\frac{1}{D_{\Sigma,p}} \times 
\\  &\begin{pmatrix} \ i\nu+\mu+D^{-1}_{0,h}|\Delta^{}_{BA}|^2(i\nu-\mu) 
& t_{\mathrm{nn}}\sum\limits_j \left[ e^{i\bm{k}\cdot \bm{R}_j}+D^{-1}_{0,h}\Delta^{}_{AB}\Delta^{*}_{BA}e^{-i\bm{k}\cdot \bm{R}_j} \right] \  
\\  \ t_{\mathrm{nn}}\sum\limits_j \left[ e^{-i\bm{k}\cdot \bm{R}_j}+D^{-1}_{0,h}\Delta^{*}_{AB}\Delta^{}_{BA}e^{i\bm{k}\cdot \bm{R}_j} \right] 
& i\nu+\mu+D^{-1}_{0,h}|\Delta^{}_{AB}|^2(i\nu-\mu)  \end{pmatrix}, 
\\
&[(G^h_{0})^{-1}-\Sigma^p]^{-1}=\frac{1}{D_{\Sigma,h}}\times 
\\  
&\begin{pmatrix} \ i\nu-\mu+D^{-1}_{0,p}|\Delta^{}_{AB}|^2(i\nu+\mu) & -t_{\mathrm{nn}}\sum\limits_j \left[ e^{i\bm{k}\cdot \bm{R}_j}-D^{-1}_{0,p}\Delta^{}_{AB}\Delta^{*}_{BA}e^{-i\bm{k}\cdot \bm{R}_j} \right] 
\\  \ -t_{\mathrm{nn}}\sum\limits_j \left[ e^{-i\bm{k}\cdot \bm{R}_j}-D^{-1}_{0,p}\Delta^{*}_{AB}\Delta^{}_{BA}e^{i\bm{k}\cdot \bm{R}_j} \right]
& i\nu-\mu+D^{-1}_{0,p}|\Delta^{}_{BA}|^2(i\nu+\mu) \end{pmatrix},
\end{split}
\end{equation}
where $D_{\Sigma,p/h}=\mathrm{Det}\left\{(G^{p/h}_{0})^{-1}-\Sigma^{h/p}\right\}$.
Plugging the self energies and normal-state Green's functions back into Eq.~(\ref{eqn:Green_SC}), and defining $2\Delta_1=\Delta_{AB}+\Delta_{BA}$ and $2i\Delta_2=\Delta_{BA}-\Delta_{AB}$, we find 
\begin{equation}
\check{G}_{}(\nu,\bm{k})=\frac{-1}{D(\nu,\bm{k})}\sum_{n,m}G_{n,m}(\nu,\bm{k})\tau_n\otimes\sigma_m,
\label{eqn:App_Green}
\end{equation}
where $t=t_{\rm nn}$ for shortness and
\begin{equation}
\begin{split}
G_{0,0}(\nu,\bm{k})=&-i\nu(\nu^2+\mu^2+t^2f_1^2+t^2f_2^2+|\Delta_1^2|+|\Delta_2^2|), \\
G_{0,1}(\nu,\bm{k})=& -2i\nu\mu tf_1,\\
G_{0,2}(\nu,\bm{k})=&-2i\nu\mu tf_2, \\
G_{0,3}(\nu,\bm{k})=& i\mu (\Delta^{}_2 \Delta_1^*-\Delta^{}_1 \Delta_2^*)
=\mu\,\frac{|\Delta_{BA}|^2-|\Delta_{AB}|^2}{2}
=-\frac{1}{2}\,\mu \, \Xi,\\
G_{1,0}(\nu,\bm{k})=&-2\mu t(f_1\mathrm{Re}\{\Delta_1\}+f_2\mathrm{Re}\{\Delta_2\}), \\
G_{1,1}(\nu,\bm{k})=&-\mathrm{Re}\{\Delta_1\}(\mu^2+\nu^2+t^2f_1^2-t^2f_2^2+|\Delta_1|^2)-2t^2f_1f_2\mathrm{Re}\{\Delta_2\}-\mathrm{Re}\{\Delta_1^*(\Delta_2)^2\}, \\
G_{1,2}(\nu,\bm{k})=&-\mathrm{Re}\{\Delta_2\}(\mu^2+\nu^2-t^2f_1^2+t^2f_2^2+|\Delta_2|^2)-2t^2f_1f_2\mathrm{Re}\{\Delta_1\}-\mathrm{Re}\{\Delta_2^*(\Delta_1)^2\}, \\
G_{1,3}(\nu,\bm{k})=&2i\nu t(f_2\mathrm{Im}\{\Delta_1\}-f_1\mathrm{Im}\{\Delta_2\}), \\
G_{2,0}(\nu,\bm{k})=&2\nu t(f_1\mathrm{Im}\{\Delta_1\}+f_2\mathrm{Im}\{\Delta_2\}), \\
G_{2,1}(\nu,\bm{k})=&\mathrm{Im}\{\Delta_1\}(\mu^2+\nu^2+t^2f_1^2-t^2f_2^2+|\Delta_1|^2)-2t^2f_1f_2\mathrm{Im}\{\Delta_2\}+\mathrm{Im}\{\Delta_1^*(\Delta_2)^2\}, \\
G_{2,2}(\nu,\bm{k})=&\mathrm{Im}\{\Delta_2\}(\mu^2+\nu^2-t^2f_1^2+t^2f_2^2+|\Delta_2|^2)+2t^2f_1f_2\mathrm{Im}\{\Delta_1\}+\mathrm{Im}\{\Delta_2^*(\Delta_1)^2\}, \\
G_{2,3}(\nu,\bm{k})=&2i\nu t(f_2\mathrm{Re}\{\Delta_1\}-f_1\mathrm{Re}\{\Delta_2\}), \\
G_{3,0}(\nu,\bm{k})=&\mu(\nu^2+\mu^2-t^2f_1^2-t^2f_2^2+|\Delta_1^2|+|\Delta_2^2|), \\
G_{3,1}(\nu,\bm{k})=&-tf_1(\nu^2-\mu^2+t^2f_1^2+t^2f_2^2+|\Delta_1^2|-|\Delta_2^2|)-tf_2(\Delta^{}_2\Delta_1^*+\Delta^{}_1\Delta_2^*), \\
G_{3,2}(\nu,\bm{k})=&-tf_2(\nu^2-\mu^2+t^2f_1^2+t^2f_2^2-|\Delta_1^2|+|\Delta_2^2|)-tf_1(\Delta^{}_2\Delta_1^*+\Delta^{}_1\Delta_2^*), \\
G_{3,3}(\nu,\bm{k})=&\nu(\Delta^{}_2\Delta_1^*-\Delta^{}_1\Delta_2^*)
=\frac{i}{2}\,\nu \, \Xi,
\end{split}
\label{eqn:All_Green}
\end{equation}
The function in the denominator of Eq.~(\ref{eqn:App_Green}) is 
\begin{equation}
D(\nu,\bm{k})=[\nu^2+E^2_{1}(\bm{k})]\,[\nu^2+E^2_{2}(\bm{k})].
\end{equation}
The zeros of the function $D(\nu,\bm{k})$, which produce poles in Green's function Eq.~(\ref{eqn:App_Green}), give the energy eigenvalues $\pm E_{1}(\bm{k})$ and $\pm E_{2}(\bm{k})$ of the BdG Hamiltonian.

\appendix{The ac Hall conductivity of a chiral superconductor}
\label{App:Matsubara}

In this Appendix, we present technical details of the evaluation of ac Hall conductivity for a chiral superconductor in Eq.~(\ref{eqn:Hall_Gj}), which is copied below for convenience of viewing
\begin{equation}
\sigma_{H}(\omega)=i\epsilon_{\alpha\beta}\frac{Te^2}{\hbar^3\omega}\lim_{i\omega_n\to\omega+i0^+}\sum_\nu\int \frac{d^2k}{(2\pi)^2}
\mathrm{Tr}_{\sigma}\sum_{q=0}^3 v_\alpha(\bm{k}) G_q(\nu,\bm{k})v_\beta(\bm{k}) G_q(\nu+\omega_n,\bm{k}).
 \label{eqn:Hall_Gj-app}
\end{equation}
The velocity matrix (\ref{dH0/dk}) for our model (\ref{eqn:Hamiltonian}) contains only the Pauli matrices $\sigma_1$ and $\sigma_2$.  Due to the antisymmetric factor $\epsilon_{\alpha\beta}$, a nonzero result is obtained in Eq.~(\ref{eqn:Hall_Gj-app}) only from the terms where one of the two velocity matrices has $\sigma_1$ and another $\sigma_2$.  Then, to obtain a nonzero trace in Eq.~(\ref{eqn:Hall_Gj-app}), one of the two Green's functions $G_q$ must have $\sigma_1$ and another $\sigma_2$, or, alternatively, $\sigma_3$ and $\sigma_0$.  The first option results in $\sigma_1^2=\sigma_0$ and $\sigma_2^2=\sigma_0$, which gives zero due to the antisymmetric factor $\epsilon_{\alpha\beta}$.  So, only the second option, where one of the Green's functions $G_q$ has $\sigma_3$ and another $\sigma_0$, can give a nonzero result.  By cyclic permutation, the unit matrix $\sigma_0$ can always be placed between the velocity matrices.  Thus, we arrive at the following result
\begin{align}
\sigma_{H}(\omega) = & -2\frac{e^2}{\hbar^3\omega}\lim_{i\omega_n\to\omega+i0^+}
\int \frac{d^2k}{(2\pi)^2} \, \Upsilon(\bm k)
\nonumber \\
 & \times T\sum_\nu \sum_{q=0}^3 \left\{ G_{q,0}(\nu) G_{q,3}(\nu+\omega_n) 
 - G_{q,3}(\nu) G_{q,0}(\nu+\omega_n) \right\},
 \label{eqn:Hall_G03}
\end{align}
where we used Eq.~(\ref{eqn:Upsilon}) and the notation of Eq.~(\ref{eqn:App_Green}).  
Substituting Eq.~(\ref{eqn:All_Green}) into Eq.~(\ref{eqn:Hall_G03}) and evaluating to sum over $q$, we find that only the terms cubic in frequency give a nonzero contribution proportional to
\begin{equation}
(\nu+\omega_n)^3-\nu^3+\nu(\nu+\omega_n)^2-\nu^2(\nu+\omega_n)
= \omega_n[\nu+(\nu+\omega_n)]^2.
\label{eqn:nu-omega}
\end{equation}
Thus, we arrive at the following expression
\begin{equation}
\sigma_{H}(\omega)=\frac{e^2}{\hbar\omega}\lim_{i\omega_n\to \omega+i0^+}\int \frac{d^2k}{(2\pi)^2}\, \Upsilon(\bm{k}) \, M(\omega_n,\bm{k}) \, \Xi(\bm{k}) \, \mu.
\label{eqn:Hall-M-Ups}
\end{equation}
Here $M(\omega_n,\bm{k})$ is the Matsubara sum
\begin{equation}
M(\omega_n,\bm{k})\equiv i\omega_n T\sum_{\nu}h(\nu,\omega_n,\bm k)
\label{eqn:M}
\end{equation}
of the function 
\begin{equation}
h(\nu,\omega_n)= \frac{(2{\nu}+\omega_n)^2}{({\nu}^2+E^2_{1})\,({\nu}^2+E^2_{2})\,[({\nu}+\omega_n)^2+E^2_{1}]\,[({\nu}+\omega_n)^2+E^2_{2}]},
\label{eqn:h}
\end{equation}
where $\bm k$ is omitted for brevity.  Note that $M(\omega_n)$ is an odd function of $\omega_n$, as follows from Eq.~(\ref{eqn:Hall_G03}).  Equations (\ref{eqn:Hall-M-Ups}), (\ref{eqn:M}) and (\ref{eqn:h}) are reminiscent of Eq.~(36) in Ref.~\refcite{Yakovenko}.

Using the poles of $h({\nu})$, $Z_n=\{\pm E_1,\pm E_2,\pm E_1-i\omega_n,\pm E_2-i\omega_n\}$, we carry out the Matsubara sum in Eq.~(\ref{eqn:M})
\begin{equation}
\begin{split}
& M(\omega_n)
=i\omega_n\sum_{Z_n} {\rm iRes}\{F(Z_n)h(-iZ_n)\} \\ 
& = \frac{i\omega_n(\omega_n+2iE_1)^2F(-E_1)}{2E_1(iE_1+iE_2)(-iE_{12})(\omega_n+2iE_1)\omega_n(\omega_n+(iE_1+iE_2))(\omega_n-iE_{12})}
\\&+
\frac{i\omega_n(\omega_n-2iE_1)^2F(E_1)}{-2E_1(-iE_1-iE_2)(iE_{12})(\omega_n-2iE_1)\omega_n(\omega_n-(iE_1+iE_2))(\omega_n+iE_{12})}
\\&+
\frac{i\omega_n(\omega_n+2iE_2)^2F(-E_2)}{2E_2(iE_1+iE_2)(iE_{12})(\omega_n+2iE_2)\omega_n(\omega_n+(iE_1+iE_2))(\omega_n+iE_{12})}
\\&+
\frac{i\omega_n(\omega_n-2iE_2)^2F(E_2)}{-2E_2(-iE_1-iE_2)(-iE_{12})(\omega_n-2iE_2)\omega_n(\omega_n-(iE_1+iE_2))(\omega_n-iE_{12})}
\\&+
\frac{i\omega_n(-\omega_n+2iE_1)^2F(-E_1-i\omega_n)}{2E_1(iE_1+iE_2)(-iE_{12})(-\omega_n+2iE_1)(-\omega_n)(-\omega_n+(iE_1+iE_2))(-\omega_n-iE_{12})}
\\&+
\frac{i\omega_n(-\omega_n-2iE_1)^2F(E_1-i\omega_n)}{-2E_1(-iE_1-iE_2)(iE_{12})(-\omega_n-2iE_1)(-\omega_n)(-\omega_n-(iE_1+iE_2))(-\omega_n+iE_{12})}
\\&+
\frac{i\omega_n(-\omega_n+2iE_2)^2F(-E_2-i\omega_n)}{2E_2(iE_1+iE_2)(iE_{12})(-\omega_n+2iE_2)(-\omega_n)(-\omega_n+(iE_1+iE_2))(-\omega_n+iE_{12})}
\\&+
\frac{i\omega_n(-\omega_n-2iE_2)^2F(E_2-i\omega_n)}{-2E_2(-iE_1-iE_2)(-iE_{12})(-\omega_n-2iE_2)(-\omega_n)(-\omega_n-(iE_1+iE_2))(-\omega_n-iE_{12})},
\end{split}
\end{equation}
where we defined $E_{12}\equiv E_2-E_1$. Using the property $F(E_i+i\omega_n)=F(E_i)$ of the Fermi function $F$ for the bosonic Matsubara frequency $\omega_n=2n\pi T$, we find
\begin{equation}
\begin{split}
M(\omega_n)
=&\frac{-i(\omega_n-2iE_1)[F(-E_1)-F(E_1)]}{2 E_1(iE_1+iE_2)(iE_{12})(\omega_n-(iE_1+iE_2))(\omega_n+iE_{12})}\\&+\frac{-i(\omega_n+2iE_1)[F(-E_1)-F(E_1)]}{2 E_1(iE_{12})(iE_1+iE_2)(\omega_n-iE_{12})(\omega_n+(iE_1+iE_2))} \\ &+\frac{i(\omega_n+2iE_2)[F(-E_2)-F(E_2)]}{2 E_2(iE_{12})(iE_2+iE_1)(\omega_n+iE_{12})(\omega_n+(iE_2+iE_1))}\\&+\frac{i(\omega_n-2iE_2)[F(-E_2)-F(E_2)]}{2 E_2(iE_2+iE_1)(iE_{12})(\omega_n-(iE_2+iE_1))(\omega_n-iE_{12})}.
\end{split}
\end{equation}

At $T=0$, we find that
\begin{equation}
\begin{split}
-iM(\omega_n)
=&\frac{-(\omega_n-2iE_1)}{2 E_1(iE_1+iE_2)(iE_{12})[\omega_n-(iE_1+iE_2)](\omega_n+iE_{12})}\\&+\frac{(\omega_n-2iE_2)}{2 E_2(iE_2+iE_1)(iE_{12})[\omega_n-(iE_2+iE_1)](\omega_n-iE_{12})} \\ &+c.c. \\
=&\frac{(\omega_n-2iE_1)E_2(\omega_n-iE_{12})-(\omega_n-2iE_2)E_1(\omega_n+iE_{12})}{2 E_1E_2(E_1+E_2)E_{12}[\omega_n-(iE_1+iE_2)](\omega_n^2+E_{12}^2)}+c.c. \\
=&\frac{\omega_n^2E_{12}-\omega_niE_{12}(E_1+E_2)-4E_1E_2E_{12}}{2 E_1E_2(E_1+E_2)(E_{12})[\omega_n-(iE_1+iE_2)](\omega_n^2+E_{12}^2)}+c.c. \\
=&\frac{\omega_n^2-i\omega_n(E_1+E_2)-4E_1E_2}{2 E_1E_2(E_1+E_2)[\omega_n-(iE_1+iE_2)](\omega_n^2+E_{12}^2)}+c.c. \\
=&\frac{(\omega_n^2-i\omega_n(E_1+E_2)-4E_1E_2)[\omega_n+i(E_1+E_2)]}{2 E_1E_2(E_1+E_2)[\omega_n^2+(E_2+E_1)^2](\omega_n^2+E_{12}^2)}+c.c. \\
=&\frac{\omega_n^3+\omega_n(E_1+E_2)^2-4\omega_nE_1E_2}{E_1E_2(E_1+E_2)[\omega_n^2+(E_2+E_1)^2](\omega_n^2+E_{12}^2)} \\
=&\frac{\omega_n(\omega_n^2+E_{12}^2)}{E_1E_2(E_1+E_2)[\omega_n^2+(E_2+E_1)^2](\omega_n^2+E_{12}^2)} \\
=&\frac{\omega_n}{E_1E_2(E_2+E_1)[\omega_n^2+(E_2+E_1)^2]}.
\end{split}
\end{equation}

\bibliographystyle{ws-ijmpb}
\bibliography{Refs}

\end{document}